\documentstyle[12pt]{article}
\evensidemargin \oddsidemargin
\catcode `\@=11
\def\numberbysection{\@addtoreset{equation}{section}
         \renewcommand{\theequation}{\thesection.\arabic{equation}}}
\def\be{\begin{equation}}
\def\ee{\end{equation}}
\newcommand{\ba}{\begin{eqnarray}}
\newcommand{\ea}{\end {eqnarray}}
\newcommand{\nn}{\nonumber}
\def\ri{\right}
\def\le{\left}

\def\a{\alpha}
\def\b{\beta}
\def\g{\gamma}

\def\d{\delta}
\def\D{\Delta}

\def\e{\eta}
\def\th{\theta}

\def\l{\lambda}
\def\L{\Lambda}

\def\s{\sigma}
\def\sp{\sigma^+}
\def\sm{\sigma^-}
\def\sx{\sigma^x}
\def\sy{\sigma^y}
\def\sz{\sigma^z}
\def\t{\tau}

\def\vfi{\varphi^{(i)}}
\def\c{\chi}
\def\O{\Omega}
\def\Oh{\widehat\Omega}

\def\ap{a^+}
\def\cp{c^+}
\def\woo{w_{0,0}}
\def\woi{w_{0,1}}
\def\wio{w_{1,0}}
\def\wii{w_{1,1}}

\def\Et{\tilde{E}}
\def\nil{\emptyset}

\def\lra{\longrightarrow}
\def\ti{\tilde}

\def\inf{\infty}
\def\ten{\otimes}

\def\pa{\partial}
\def\half{\frac{1}{2}}
\def\quart{\frac{1}{4}}
\numberbysection
\begin{document}
% TITLE-PAGE
%
\pagestyle{empty}
\vspace* {10mm}
\renewcommand{\thefootnote}{\fnsymbol{footnote}}
\begin{center}
\large
   {\bf FINITE-SIZE SCALING STUDIES}\\[4mm]
   {\bf OF REACTION-DIFFUSION SYSTEMS}\\[4mm]
   {\bf Part I: Periodic Boundary Conditions}\\[1cm]
\end{center}
\begin{center}
\normalsize
   Klaus Krebs, Markus Pfannm\"uller and Birgit Wehefritz \\[5mm]
   {\it Universit\"{a}t Bonn,
   Physikalisches Institut \\ Nu\ss allee 12,
   D-53115 Bonn, Germany}
\\[2cm]
{\bf Abstract}
\end{center}
\renewcommand{\thefootnote}{\arabic{footnote}}
\addtocounter{footnote}{-1}
%
%
%Abstract
\small

The finite-size scaling function and the leading corrections for the
single species 1D coagulation model $(A + A \rightarrow A)$ and the
annihilation model $(A + A \rightarrow \emptyset)$ are calculated.
The scaling functions are universal and independent of the initial conditions
but do depend on the boundary conditions. A similarity transformation
between the two models is derived and used to connect the corresponding
scaling functions.
\\
\rule{5cm}{0.2mm}
\begin{flushleft}
\parbox[t]{3.5cm}{\bf Key words:}
\parbox[t]{12.5cm}{Reaction-diffusion systems, finite-size scaling,
                   non-equilibrium statistical mechanics,
                   coagulation model, annihilation model}
\\[2mm]
\parbox[t]{3.5cm}{\bf PACS numbers:}
\parbox[t]{12.5cm}{05.40.+j, 05.70.Ln, 82.20.Mj}
\end{flushleft}
\normalsize
\vspace{3cm}
\begin{flushleft}
BONN HE-93-51\\
cond-mat/9402017\\
Bonn University\\
December 1993\\
ISSN-0172-8733
\end{flushleft}
\thispagestyle{empty}
\mbox{}
\newpage
\setcounter{page}{1}
\pagestyle{plain}
%-------------------------------
%
% THE INTRODUCTION
%
\section {Introduction}
\label {sec:intro}
\hspace{\parindent}Since
Smoluchowski \cite{Smoluchowski}
demonstrated that the macroscopic phenomenon of
diffusion can be explained on the microscopic scale by Brownian
motion of particles, reaction-diffusion systems are
a field of intense research in non-equilibrium statistical mechanics.
There are numerous applications in physical chemistry as well as
in physics and biology: propagation of
phonons in solids, biochemical reactions or catalysts, to name only a few.
Even in the simplest models exact results are very scarce because
analytic calculations are often difficult.
Monte-Carlo simulations and mean field calculations or
truncation schemes for an infinite hierarchy of $n$-point functions
\cite{Kuzovkov} have been
the most promising methods for a long time.

Reaction-diffusion systems can be described by lattice models.
The dynamics are given by a master equation for the
probability $P(\{\b\},t)$ to have the configuration $\{\b\}$ realized at
time $t$ \cite{Kadanoff,Grassberger}.
Concerning one-dimensional systems, the mapping of the whole problem
onto a quantum chain \cite{Alcaraz} is a big progress.
The result is the euclidean Schr\"odinger equation
\be
 \frac{\pa}{\pa t}P(\{\b\},t) = - HP(\{\b\},t) \;.
\ee
This formulation has various advantages. On the one hand, using a similarity
transformation between the Hamiltonians, one can prove
the equivalence of different models. On the other hand, methods which have been
restricted to equilibrium statistical mechanics, can be applied to a new
class of problems.
In this context $q$-deformed quantum groups and Hecke
algebras, which have mainly been of mathematical interest up to now,
arise naturally \cite{Alcaraz3}. It is possible as well to use Bethe
Ansatz techniques in non-equilibrium problems \cite{Gwa1,Gwa2}.

Finite-size scaling was originally proposed in equilibrium statistical
mechanics.
There one is interested in phase transitions, i.e. the singularities of
thermodynamic quantities at the critical temperature $T_c$.
These singularities become smooth if the system has a finite size.
The idea is to extract critical exponents of an
infinite system by studying how thermodynamic quantities vary with
the size of the finite system \cite{Barber,Malte}.

The key hypothesis of finite-size scaling in thermodynamics
is that in
the vicinity of the bulk critical temperature $T_c$ there should be only one
relevant length describing the smoothing and shifting of thermodynamic
singularities. Therefore the behaviour of the finite system should be
determined by the scaled variable
\be
x\;=\;\frac{L}{\xi(T)}
\ee
where $\xi(T)$ is the bulk correlation length and $L$ is the characteristic
length scale of the system.
Let $P(T)$ denote a thermodynamic quantity which in
the infinite system can be written as
\be
P_{\infty}(T)\sim C_{\infty} \le(\frac{T-T_c}{T_c}\ri)^{-\rho}.
\ee
In the
finite-size scaling regime ($L\rightarrow\infty$, $T\rightarrow T_c$ but
$x$ remaining finite) it is supposed to behave like
\be
P_{L}(T)\sim \;L^{\omega}Q(x)\;.
\ee
Consistency between the two limits requires a relation between
the two critical exponents $\omega$
and $\rho$ such that the variation of the thermodynamic
quantities due to finite-size effects is determined by the bulk
critical exponents.

Quantum chains are another application of finite-size scaling.
Here critical exponents are extracted from the
spectrum of the Hamiltonian.

It has been proposed in \cite{Alcaraz} to use
finite-size scaling in non-equilibrium chemical models to treat, for example,
the concentration of particles per site $\frac{<n>}{L}$. Another example are
the studies of
deposition-evaporation systems \cite{Stinch} where the scaling properties of
a special kind of correlation function are discussed.

The scaling limit we consider in the investigation of reaction-diffusion
systems differs from the one considered in equilibrium problems,
namely we keep $z=\frac{4Dt}{L^2}$ constant while $t$ and $L$ go to
infinity. Here $D$ denotes the
diffusion constant.
This limit has its origin in the consideration
of pure diffusion processes described by $D \Delta c(t)=\partial_{t}c(t)$.
The average of the square of the distance travelled by a particle in time $t$
is then given by $<(x_t-x_0)^2>=2Dt$. Thus
$z=\frac{4Dt}{L^2}$ appears naturally as the square of the
``correlation length'' devided by $L^2$.
The additional factor of two is for later convenience.
Applying the scaling hypothesis to the concentration of particles gives:
\be
c(z,L) = L^x\:[F_{0}(z)+L^{-y}\:F(z)+\cdots]\;.
\label{eq:scal1}
\ee
Here $F_{0}$ denotes the scaling function, $x$ the scaling
exponent and $L^{-y}\:F(z)$ is the leading correction term. $F_0$ and the
correction function $F$ depend only on the scaling variable $z$.
The scaling exponent $x$ and the critical exponent $\a$, that describes the
large-time behaviour of an infinite system, i.e. $c(t) \simeq t^\a$, are
connected by the scaling relation:
\be
\a=\frac{x}{2}\;.
\ee
Similar to equilibrium statistical mechanics the existence of finite-size
scaling relations allows the numerical determination of critical exponents
($\a$ in the present case) from finite lattices.

The question of universality
arises immediately. Do the
exponents and scaling functions depend on the details of the model and
the initial conditions?

We will concentrate on systems with a single type of particle
(denoted by $A$) and the following reactions:
\begin{itemize}
\item diffusion:
\[ A + \nil \rightleftharpoons \nil + A \;,\]
\item coagulation and decoagulation:
\[ A + A \rightleftharpoons A \;,\]
\item annihilation and creation:
\[ A + A \rightleftharpoons \nil\;,\]
\end{itemize}
where $\nil$ denotes a state without any particles.

Our aim was to present the results we obtained as clearly as
possible. Thus we decided to publish a series of three papers:
the purpose of the present one is to briefly review the
master equation approach
to reaction-diffusion systems and to present connections between
different
models that are valid for all boundary conditions. Using periodic boundaries
we derive exact results for the particle concentration in the coagulation
and the annihilation model. The scaling and the long-time behaviour
are studied.

Since the calculations for open boundaries require different techniques
we turn to this case in a second paper \cite{Pap2}.
There we will present the exact solution of the coagulation-decoagulation
model.  Again, the applicability of scaling arguments will be investigated.
The scaling exponent is expected to be universal.
However it is important to know the scaling functions and
the corrections, for example to decide whether numerical diagonalisations
for small lattices are best performed with open or with periodic boundaries.

The third article \cite{Pap3} will be devoted to numerical techniques.
The question whether extrapolations from small lattices can be used to
determine scaling exponents and functions will be discussed.
We will use Monte Carlo techniques to study weakly correlated
initial conditions and the case of reaction rates where analytical methods
cannot be applied easily.

This paper is organized as follows:
In Section \ref{sec:master}, we present the general framework we used to
study reaction-diffusion processes.
In Section \ref{sec:ham}, we define our models and give
the Hamiltonians describing them. We find that
for special ratios of the reaction  rates the spectrum
of the coagulation-decoagulation model can be expressed in
terms of free fermions. We present these calculations in Section
\ref{sec:diag}.
We construct an explicit similarity transformation mapping the
Hamiltonian of the coagulation model onto the one corresponding to
the annihilation model in Section \ref{sec:connex}. As an application,
we give the relations between all $n$-point functions of the two models.
Therefore we can finally conclude that both systems are equivalent.

In Section \ref{sec:holes}, an exact expression for the
concentration $c(t, L)$ of particles in the coagulation-decoagulation
model
is calculated for arbitrary lattice lengths $L$.
In Section \ref{sec:fin_size}
we investigate the scaling behaviour of the concentration in the case of
a vanishing decoagulation rate.
Exponents,
the scaling function and the corrections for finite lattices
are determined.
The influence of the initial occupation probabiltity in uncorrelated initial
states on the exponents and the scaling functions is studied.
Finally we discuss the relation between the thermodynamic and the scaling
limit.
The scaling function
is derived from the continuous version of the model in
Appendix~\ref{sec:Cont}.

Section \ref{sec:anni} is devoted to the annihilation model. Using
the similarity transformation we derive exact results for the concentration
from the expressions obtained for the coagulation model. We calculate
the scaling exponents and functions and the finite lattice corrections.

We close with a discussion of our results. Open
questions and possible directions of further research are outlined.
%---------------------------------------------------
%
% CHAPTER: MASTER EQUATION and QUANTUM CHAINS
%
%
\section {Master Equation and Quantum Chains}
\label{sec:master}
\hspace{\parindent}The
models studied in this paper are defined on an one-dimensional lattice of
length $L$. To each site $i$ we attach a variable $\b_{i}$.
$\b_{i}=0$ corresponds to a vacancy, $\b_{i}=1$ to a particle
of type $A$. Throughout this paper we will always impose periodic boundary
conditions.

Let $P(\lbrace\b\rbrace;t)$ be the probability to find the system at time $t$
in the configuration $\lbrace\b\rbrace$. The dynamics of the system are then
determined by the following master equation \cite{Alcaraz} describing the
time evolution of the probability distribution $P(\lbrace\b\rbrace;t)\:$:

\begin {eqnarray}
\frac{ \partial}{\partial t}P(\lbrace\b\rbrace;t)
     & = & \sum_{k=1}^L\biggl[
             -w_{0,0}(\b_{k},\b_{k+1}) P(\b_{1}, \ldots,\b_{L};t)\nn\\*
     &   & +\sum_{l,m=0}^{1}\!\!\!^\prime\, w_{l,m}(\b_{k},\b_{k+1})
                P(\b_{1},\ldots,\b_{k}+l,\b_{k+1}+m,\ldots,\b_{L};t)\biggr]\;.
\label {eqn:master}
\end {eqnarray}

A prime is used to indicate that in a sum the pair $l=m=0$ is excluded.
All arithmetic operations on the $\b_i$ are performed modulo 2.

The probability that a state $(\g,\d)$ on two adjacent sites will
change into the state $(\a,\b)$ after one unit of time is given by:
\be
w_{\g-\a,\d-\b}(\a,\b)\;;\;(\a,\b)\not=(\g,\d).
\ee
The rates $w_{0,0}(\a,\b)$ are related to the probability that in one unit of
time the state $(\a,\b)$ remains unchanged. From the conservation of
probability we have:
\be
w_{0,0}(\a,\b)=\sum_{r,s}\,^\prime\, w_{r,s}(\a-r,\b-s)\;,
\ee
where again $r=s=0$ is excluded. Clearly all rates have to be non-negative
and real.

In order to rewrite the master equation (\ref{eqn:master}) in the form of a
Schr\"odinger equation with a quantum chain Hamiltonian,
we attach a two-dimensional vector space $V_i$ to
each site $i$. The vector $1 \choose 0$ corresponds to a vacancy,
$0 \choose 1$ to a particle.
We define the $2 \times 2$ matrices $E^{kl}$
with entries $(E^{kl})_{nm}=\d_{k,n}\d_{l,m}$ and the matrix $F$:
\be
F = \le(
    \begin{array}{cc}
    0 & 1 \\
    1 & 0
    \end{array}
    \ri)\;;\;F^2=1\;.
\ee
Using these notations we can define a quantum chain Hamiltonian:
$\widetilde H$
\ba
\widetilde H    & = & \sum_{i=1}^L \widetilde H_i \;,
\label{eqn:H}\\
\widetilde H_i  & = & \widetilde U_i-\widetilde T_i\;.
\ea
$\widetilde H$ operates on the $L$-fold tensor product $V_1\ten\cdots\ten V_L$.
The $\widetilde H_i$ operate locally on $V_i\ten V_{i+1}$.
The kinetic energy part $\widetilde T_i$ is non-diagonal and the
potential energy part $\widetilde U_i$ is diagonal:
\ba
\widetilde T_i & = & \sum_{l,m=0}^1\!\!\!^\prime\sum_{\a,\b=0}^1
              w_{l,m}(\a,\b)\Bigl(E^{\a\a}F^l\Bigr)_i
                              \ten\Bigl(E^{\b\b}F^m\Bigr)_{i+1}\;\;,
\label{eqn:Ti}\\
\widetilde U_i & = & \sum_{\a,\b=0}^1\woo(\a,\b)E^{\a\a}_i
                                 \ten E^{\b\b}_{i+1}\;.
\label{eqn:Ui}
\ea
The master equation (\ref{eqn:master}) can now be replaced by the euclidean
Schr\"odinger equation:
\be
\frac{\pa}{\pa t}|P\rangle = -\widetilde H|P\rangle\;.
\label{eqn:schro2}
\ee
At this point a comment on the states $|P\rangle$ might be helpful.
We choose an orthonormal basis which we are going to call the spin basis:
\cite{Kadanoff}
\be
|\{\b\}\rangle = |\b_1,\ldots,\b_L\rangle
\;;\hspace{1cm} \langle\{\b\}|\{\b'\}\rangle = \d_{\{\b\},\{\b'\}}\;,
\ee
and define the ket state:
\be
|P\rangle = \sum_{\{\b\}}P(\{\b\};t)|\{\b\}\rangle\;.
\ee
The wave function already represents a probability.
Thus, in general the calculation of expectation values differs from ordinary
quantum mechanics. Take an observable $X$ (for example
the concentration of particles). Instead of $<X>(t) = \langle P|X|P\rangle$
as in quantum mechanics, the average is computed as:
\ba
<X>(t) & = & \sum_{\{\b\}}X(\{\b\})P(\{\b\};t)\nn\\
       & = & \langle 0|X|P\rangle\nn\\
       & = & \langle 0|Xe^{-\widetilde H\;t}|P_0\rangle
\label{eq:exval1}
\ea
with the bra ground state:
\be
\langle 0| = \sum_{\{\b\}}\langle\{\b\}|
\label{eq:brags}
\ee
and the initial ket state:
\be
|P_0\rangle = \sum_{\{\b\}}P(\{\b\};0)|\{\b\}\rangle\;.
\ee
The calculation of expectation values can often be simplified.
Let $X$ be a local operator, i.e. the matrix
elements depend non-trivially only on the configuration of a subset
$\{i_1,\ldots,i_k\}$ of sites:
\be
\langle\b_1,\ldots,\b_L|X|\b'_1,\ldots,\b'_L\rangle =
\langle\b_{i_1},\ldots,\b_{i_k}|X|\b'_{i_1},\ldots,\b'_{i_k}\rangle
\prod_{j \not= i_1,\ldots,i_k} \d_{\b_j,\b'_j}
\;.
\ee
On the right hand side
$\langle\b_{i_1},\ldots,\b_{i_k}|X|\b'_{i_1},\ldots,\b'_{i_k}\rangle$
contains only the non-trivial dependencies.
We define a new operator $\overline{X}$ with matrix elements:
\be
\langle\b_1,\ldots,\b_L|\overline{X}|\b'_1,\ldots,\b'_L\rangle =
\prod_{j = 1}^L \d_{\b_j,\b'_j}
\sum_{\b''_{i_1},\ldots,\b''_{i_k}}
\langle\b''_{i_1},\ldots,\b''_{i_k}|X|\b'_{i_1},\ldots,\b'_{i_k}\rangle
\;.
\label{eq:op_diag}
\ee
Turning to a matrix representation in the spin basis, it is clear that
$\overline{X}$ is diagonal. Each diagonal element is obtained from the matrix
representing $X$ by summing up all entries of the corresponding column.
Now we can state a useful identity:
\be
\langle 0|X|\{\b\}\rangle=\langle 0|\overline{X}|\{\b\}\rangle\;.
\label{eq:exval2}
\ee
The proof is straightforward:
\ba
\langle 0|X|\{\b'\}\rangle & = & \sum_{\{\b\}}\biggl(
\langle\b_{i_1},\ldots,\b_{i_k}|X|\b'_{i_1},\ldots,\b'_{i_k}\rangle
\prod_{j \not= i_1,\ldots,i_k} \d_{\b_j,\b'_j}\biggr)\nn\\
 & = & \sum_{\b_{i_1},\ldots,\b_{i_k}}
\langle\b_{i_1},\ldots,\b_{i_k}|X|\b'_{i_1},\ldots,\b'_{i_k}\rangle\nn\\
 & = & \sum_{\{\b''\}}\biggl(\prod_{j=1}^L\d_{\b''_j,\b'_j}
\sum_{\b_{i_1},\ldots,\b_{i_k}}
\langle\b_{i_1},\ldots,\b_{i_k}|X|\b'_{i_1},\ldots,\b'_{i_k}\rangle
\biggr)\nn\\
 & = & \sum_{\{\b''\}}\langle\{\b''\}|\overline{X}|\{\b'\}\rangle\nn\\
 & = & \langle 0|\overline{X}|\{\b'\}\rangle\;.
\ea
Next consider two local operators $X$ and $Y$. $X$ operates at sites
$\{i_1,\ldots,i_k\}$, $Y$ at sites $\{i'_1,\ldots,i'_{k'}\}$. If the sets
$\{i_1,\ldots,i_k\}$ and $\{i'_1,\ldots,i'_{k'}\}$ are disjoint, we have
$\overline{XY}=\overline{X} \:\overline{Y}$.

Hamiltonians describing reaction-diffusion systems can be non-hermitian.
Following \cite{Alcaraz} we introduce a special kind of similarity
transformation. In several cases it will allow us to transform
non-hermitian Hamiltonians into hermitian ones:
\be
P(\lbrace\b\rbrace;t) = \Phi(\{\b\})\Psi(\{\b\};t)\;.
\label{eqn:simtr}
\ee
This transformation simply rescales the probability $P(\{\b\},t)$ of a
configuration $\{\b\}$ by a factor of $\Phi(\{\b\})^{-1}$. It is therefore
diagonal in the spin basis. One should keep in mind that the interpretation
of $\Psi(\{\b\},t)$ as probability is no longer valid.
We take $\Phi(\{\b\})$ as a local transformation of the form:
\be
\Phi(\{\b\})=\prod_{i=1}^L h^{(i)}(\b_i)\;.
\ee
Obviously all $h^{(i)}$ have to be non-zero to keep the transformation
$\Phi$ invertible. From Eq. (\ref{eqn:master}) it folllows, that $\Psi$ is a
solution of the new master equation:
\ba
\frac{ \partial}{\partial t}\Psi(\lbrace\b\rbrace;t)
     & = & \sum_{i=1}^L\biggl[
             -w_{0,0}(\b_{i},\b_{i+1}) \Psi(\b_{1}, \ldots,\b_{L};t)\nn\\
     &   & +\sum_{l,m=0}^{1}\!\!\!^\prime\, W_{l,m}^{(i)}(\b_{i},\b_{i+1})
             \Psi(\b_{1},\ldots,\b_{i}+l,\b_{i+1}+m,\ldots,\b_{L};t)\biggr]
\label{eqn:master2}
\ea
with:
\be
W_{l,m}^{(i)}(\a,\b) = \frac{\vfi(\a+l,\b+m)}{\vfi(\a,\b)}
                       w_{l,m}(\a,\b)
\ee
and:
\be
\vfi(\a,\b) = h^{(i)}(\a)h^{(i+1)}(\b)\;.
\label{eq:simtrafi}
\ee
Using $W^{(i)}(\a,\b)$ instead of $w(\a,\b)$ in Eq. (\ref{eqn:Ti})
we can rewrite Eq. (\ref{eqn:master2}) in the form of an euclidean
Schr\"odinger equation as well:
\be
\frac{\pa}{\pa t}|\Psi\rangle = -H|\Psi\rangle\;.
\label{eqn:schro}
\ee
Now expectation values are calculated from:
\ba
<X>(t) & = & \sum_{\{\b\}}X(\{\b\})\Phi(\{\b\})\Psi(\{\b\};t)\nn\\
       & = & \langle 0|X\Phi e^{-H\;t}|\Psi_0\rangle
\label{eq:exval3}
\ea
with:
\be
|\Psi_0\rangle = \sum_{\{\b\}}\Phi^{-1}(\{\b\})|\{\b\}\rangle\;.
\ee
Similarity transformations which are not diagonal will be used
in Section \ref{sec:ham}.

After this outline of the general framework, we can proceed and derive
the Hamiltonians for two-state models.
%-----------------------------------------------------------
%
%CHAPTER: THE VARIOUS HAMILTONIANS
%
\section{Two-State Hamiltonians}
\label{sec:ham}
\hspace{\parindent}In
this section the Hamiltonians for the two systems considered in this work
are presented. All Hamiltonians are given for periodic boundary conditions.
We always consider the left-right symmetric cases (i.e. $w_{\a,\b}(\g,\d)=
w_{\b,\a}(\d,\g)$) and take units of time such that we have
\begin{itemize}
\item diffusion with rate $\wii(1,0)=1$
\[
\nil + A \lra A + \nil \; .
\]
\end{itemize}
\subsection{The Coagulation-Decoagulation Model}
\label{sec:hamcd}
\hspace{\parindent}Consider
a system in which besides diffusion the following two
reactions may take place:
\begin{itemize}
\item coagulation with rate $\wio(0,1)$
\[
A + A \lra \nil + A \; ,
\]
\item decoagulation with rate $\wio(1,1)$
\[
\nil + A \lra A + A \; .
\]
\end{itemize}
Then the Hamiltonian becomes:
\ba
H   & = & \half\sum_{i=1}^{L}\biggl[
          -(\sx_i\sx_{i+1}+\sy_i\sy_{i+1})\nn\\
    &   & -\wio(0,1)\le[\frac{\vfi(1,1)}{\vfi(0,1)}\sp_i(1-\sz_{i+1})
                       +\frac{\vfi(1,1)}{\vfi(1,0)}(1-\sz_i)\sp_{i+1}\ri]\nn\\
    &   & -\wio(1,1)\le[\frac{\vfi(0,1)}{\vfi(1,1)}\sm_i(1-\sz_{i+1})
                       +\frac{\vfi(1,0)}{\vfi(1,1)}(1-\sz_i)\sm_{i+1}\ri]\nn\\
    &   & +\biggl(\frac{\woo(1,1)}{2}-\woo(1,0)\biggr)\sz_i\sz_{i+1}\nn\\
    &   & -\frac{\woo(1,1)}{2}(\sz_i+\sz_{i+1})
          +\woo(1,0)+\half\woo(1,1)\biggr]\;.
\label{eq:HCD1}
\ea
{}From the conservation of probability we have:
\ba
\woo(1,1) & = & 2\wio(0,1)\\
\woo(1,0) & = & \wii(0,1)+\wio(1,1)\; = \;1+\wio(1,1)\;.
\ea
In order to obtain a hermitian Hamiltonian, we specify the transformation
(\ref{eq:simtrafi}):
\be
h^{(i)}(1)=\sqrt{\frac{\wio(1,1)}{\wio(0,1)}} \;, \hspace{1cm}
h^{(i)}(0)=1\;;\hspace{1cm} i = 1,\ldots,L\;.
\label {eq:traf1}
\ee
Then the Hamiltonian can be written in two parts
$H = H_0 + H_1$, where
$H_0$ is the Hamiltonian of the well known XXZ-chain with an external magnetic
field in $z$-direction \cite{XXZ}:
\ba
H_0 & = & -\half\sum_{i=1}^{L}
           \Bigl[\sx_i\sx_{i+1}+\sy_i\sy_{i+1}+\D\sz_i\sz_{i+1}
          +(1-\D')(\sz_i+\sz_{i+1})+2\D'-\D-2\Bigr]
\label{eq:H0CD2}\\
H_1 & = & -\quart\sqrt{(\D-\D')(1-\D')}\sum_{i=1}^{L}
           [\sx_i(1-\sz_{i+1})+(1-\sz_i)\sx_{i+1}]
 \label{eq:H1CD2}
\ea
where:
\be
\D' =  1-\wio(0,1)\;,\hspace{1cm}
\D  = \D'+\wio(1,1)\;.
\ee
If the decoagulation rate $\wio(1,1)$ vanishes the transformation
(\ref{eq:traf1}) becomes singular.
But as we will see later even in that case
$H$ gives the correct spectrum. This is expected, because for finite $L$
the spectrum depends continuously on the parameters $\D$ and $\D'$.
$H_0$ determines the spectrum and $H_1$ is replaced by a non-hermitian
operator that does not affect the eigenvalues.

The next step is to perform a rotation around the $\sy$-axis in the space
of Pauli matrices:
\be
\sy = \ti{\s}^y ;\hspace{1cm}
\sx = \frac{\ti{\s}^x+\sqrt{\d}\ti{\s}^z}{\sqrt{1+\d}} ;\hspace{1cm}
\sz = \frac{\ti{\s}^z-\sqrt{\d}\ti{\s}^x}{\sqrt{1+\d}} \;,
\label{eq:traf2}
\ee
where:
\be
\d = \frac{\wio(1,1)}{\wio(0,1)}\;.
\ee
Introducing the notation $\e=\sqrt{1+\wio(1,1)}$ and dropping all
tildes, the Hamiltonian can be written as:
\ba
H = H_0 + H_1 & = &  -\half\e\sum_{i=1}^{L}\biggl[
            \e\sx_i\sx_{i+1}+\frac{1}{\e}\sy_i\sy_{i+1}
           +\frac{\D'}{\e}\sz_i\sz_{i+1}\nn\\
  &  &    +\frac{\sqrt{(1-\D')(\e^2-\D')}}{\e}(\sz_i+\sz_{i+1})
	  -\e-\frac{1-\D'}{\e}\biggr]\; .
\label{eq:HCD3}
\ea
For $\D'=0$ we find the Hamiltonian of the XY-chain in an external magnetic
field \cite{Hinrichsen}.

In the case of a vanishing decoagulation rate $\wio(1,1)$, i.e.
$\D = \D'$,
we take all the $\vfi$ in Eq. (\ref{eq:HCD1}) equal to one. The Hamiltonian
then
becomes $H=H_0+H_1$:
\ba
H_0 & = & -\half\sum_{i=1}^{L}
           \Bigl[\sx_i\sx_{i+1}+\sy_i\sy_{i+1}+\D'\sz_i\sz_{i+1}
          +(1-\D')(\sz_i+\sz_{i+1})+\D'-2\Bigr]\nn\\
H_1 & = & -\half(1-\D')\sum_{i=1}^{L}
          [\sp_i(1-\sz_{i+1})+(1-\sz_i)\sp_{i+1}] \;,
\label{eq:HC1}
\ea
with:
\be
\D'=1-\wio(0,1) \;.
\ee
As $H_0$ conserves the total spin $\sum \sz_i$, it can be brought into
block-diagonal form, each block corresponding to a fixed number of spins up.
If we arrange the blocks in a decreasing order, it is clear that $H_1$
leads
only to blocks above the diagonal because it increases the total
spin by two. Thus the spectrum of $H$
is determined by $H_0$ alone. The eigenvectors are of course changed by the
presence of $H_1$. If the spectrum is degenerate and $H_1$ couples two
eigenvectors belonging to the same eigenvalue the degeneracies may be
changed and $H$ might not be diagonalisable at all (this is not the fact
in our case). For a detailed discussion see Appendix A of \cite{Alcaraz}.
Observe that $H_0$ has exactly the form of the Hamiltonian (\ref{eq:H0CD2})
because we now have $\D=\D'$.
\subsection{The Annihilation-Creation Model}
\label{sec:hamac}
\hspace{\parindent}In
this system the possible reactions are diffusion
(rate $\wii(1,0)=1$) and
\begin{itemize}
\item annihilation with rate $\wii(0,0)$
\[
A + A \lra \nil + \nil \; ,
\]
\item creation with rate $\wii(1,1)$
\[
\nil + \nil \lra A + A \; .
\]
\end{itemize}
This model has been studied previously by \cite{Lushnikov,Amar,Privman1}
using various methods.
The Hamiltonian describing the model is given by:
\ba
H & = & \sum_{i=1}^{L}\Biggl[
        -\half(\sx_i\sx_{i+1}+\sy_i\sy_{i+1})\nn\\
  &   &	 +\quart[\woo(0,0)+\woo(1,1)-2\woo(1,0)]\sz_i\sz_{i+1}\nn\\
  &   &  -\frac{\vfi(1,1)}{\vfi(0,0)}\wii(0,0)\sp_i\sp_{i+1}
         -\frac{\vfi(0,0)}{\vfi(1,1)}\wii(1,1)\sm_i\sm_{i+1}\nn\\
  &   &	 +\quart[\woo(0,0)-\woo(1,1)](\sz_i+\sz_{i+1})\nn\\
  &   &  +\quart[\woo(0,0)+2\woo(1,0)+\woo(1,1)]
         \Biggr]\;.
\label{eq:HAC1}
\ea
{}From the conservation of probability we have:
\ba
\woo(1,1) & = & \wii(0,0)\;,\\
\woo(1,0) & = & \wii(0,1)\;=\;1\;,\\
\woo(0,0) & = & \wii(1,1)\; .
\ea
In this case we make a transformation (\ref{eq:simtrafi}) of the form:
\be
h^{(i)}(1)=\le(\frac{\wii(1,1)}{\wii(0,0)}\ri)^\quart\;, \hspace{1cm}
h^{(i)}(0)=1\;; \hspace{1cm} i = 1,\ldots,L \;.
\label{eq:antraf}
\ee
Once again we get a hermitian Hamiltonian $H = H_0 + H_1$:
\ba
H_0 & = & -\half\sum_{i=1}^{L}
           \Bigl[\sx_i\sx_{i+1}+\sy_i\sy_{i+1}+\D\sz_i\sz_{i+1}
          +(1-\D')(\sz_i+\sz_{i+1})+\D-2\Bigr]
          \label{eq:H0AC2}\\
H_1 & = & -\half\sum_{i=1}^{L}
           \sqrt{(\D'-\D)(2-\D-\D')}(\sx_i\sx_{i+1}-\sy_i\sy_{i+1})
          \label{eq:H1AC2}
\ea
with:
\be
   \D' = 1-\frac{\wii(0,0)-\wii(1,1)}{2}\;,\hspace{1cm}
   \D  = \D'-\wii(1,1)\;.
\ee
As before the transformation (\ref{eq:antraf}) becomes singular
when the backward reaction,
which is now creation of pairs (rate $\wii(1,1)$), does not occur.
Using arguments similar to the ones used in the coagulation-decoagulation case,
it can be seen that the Hamiltonian obtained by taking $\wii(1,1)=0$ after
the transformation still gives the correct eigenvalues.
To get $H$ in a form similar to Eq. (\ref{eq:HCD3}) we introduce the notation:
\be
\e = \le(\frac{1+\sqrt{\wii(1,1)\wii(0,0)}}
              {1-\sqrt{\wii(1,1)\wii(0,0)}}\ri)^\half\;.
\ee
Then the Hamiltonian can be written as:
\ba
H & = & -\half\sqrt{1-\wii(1,1)\wii(0,0)}\sum_{i=1}^{L}
           \biggl[\e\sx_i\sx_{i+1}+\frac{1}{\e}\sy_i\sy_{i+1}\nn\\
  &   & +\frac{1}{\sqrt{1-\wii(1,1)\wii(0,0)}}
             \bigl(\D\sz_i\sz_{i+1}
              +(1-\D')(\sz_i+\sz_{i+1})
     	  +\D-2\bigr)\biggr]\; .
\label{eq:HAC3}
\ea
After this preliminary work we will have a closer look at the spectra in
the following section and investigate the possibilities for similarity
transformations between the two models in Section \ref{sec:connex}.
%
%---------------------------------------------
%
\section{Spectrum of the Coagulation-Decoagulation Model:
\protect\\
Free Fermion Approach}
\label{sec:diag}
\hspace{\parindent}The
spectrum of the Hamiltonian (\ref{eq:HCD3}) describing
the coagulation-decoagulation model can be obtained
in terms of free fermions if we use a special choice for the rates:
\ba
\mbox{coagulation-rate } \wio(0,1) & = & 1 \;\;\mbox{( = diffusion-rate)}\nn\\
\mbox{decoagulation-rate } \wio(1,1) & \not= & 0 \;.\nn
\ea
Thinking of a model where the particles diffuse, and react with a certain
probability when two of them meet,
this choice of rates corresponds to a model where the coagulation is
instantaneous.
In the continuum limit, this model has been studied in
\cite{Burschka,ben-Avraham2}.
With:
\be
\D' = 0\;,\hspace{1cm}
\e  = \sqrt{1+\wio(1,1)}\;,
\ee
the Hamiltonian (\ref{eq:HCD3}) becomes:
\be
H = -\half\e\sum_{i=1}^{L}\biggl[
    \e\sx_i\sx_{i+1}+\frac{1}{\e}\sy_i\sy_{i+1}
    +\sz_i+\sz_{i+1}-\e-\frac{1}{\e}\biggr]\, .
\label{eq:HC=DF}
\ee
The corresponding quantum-chain is the XY-chain, which is known to be
integrable in terms of free fermions \cite{Barouch,McCoy,Suzuki}.

Remember that we had to use two similarity transformations (\ref{eq:traf1}) and
(\ref{eq:traf2}) and that this Hamiltonian describes the time evolution of
the model correctly only if the decoagulation rate $\wio(1,1)$ is non-zero.
Nevertheless, for $\wio(1,1)=0$, we still get the correct spectrum.
Now we present the diagonalisation of the
quantum chain given by Eq. (\ref{eq:HC=DF}) with periodic boundary
conditions using the method of fermionisation \cite{Barouch,Lieb}.

As a first step we perform a rotation around the $\sy$ axis in the space of
Pauli matrices; this transformation does not change the spectrum of $H$:
\be
\sx = -\tilde{\sx}\;,\hspace{1cm}
\sy = \tilde{\sy}\;,\hspace{1cm}
\sz = -\tilde{\sz}\; .
\ee
Dropping all tildes, we obtain for the Hamiltonian:
\be
H = \: -\frac{\e}{2}\sum_{i=1}^L\Biggl[\e\:\sx_i\:\sx_{i+1}+\frac{1}{\e}\:
\sy_i\sy_{i+1}-2\:\sz_i-\e-\frac{1}{\e} \Biggr]\;.
\ee
Denoting:
\be
\sigma^{\pm}\:=\:\half(\sx\:\pm i\;\sy)\,,
\ee
the Hamiltonian can be rewritten as:
\ba
H &=& -\frac{\e}{2}\sum_{i=1}^L\Biggl[
                    \biggl(\e-\frac{1}{\e}\biggr)\:\sp_i\sp_{i+1}+
                    \biggl(\e-\frac{1}{\e}\biggr)\:\sm_i\sm_{i+1}\nn\\
  & &   +\biggl(\e+\frac{1}{\e}\biggr)\:\sp_i\sm_{i+1}+
                    \biggl(\e+\frac{1}{\e}\biggr)\:\sm_i\sp_{i+1}
                         -\:4\:\sp_i\sm_i\,+\,2-\e-\frac{1}{\e}  \Biggr]\;.
\label{eqnarray:Hs}
\ea
Now one can introduce fermionic operators $a_n$ and $a_n^+$, which obey the
anti-commutation relations:
\be
\{a_i^+,a_j\}\;=\;\delta_{ij}\;; \hspace{1cm}
\{a_i^+,a_j^+\}\;=\;\{a_i,a_j\}\;=\;0
\ee
by using a Jordan-Wigner transformation:
\be
\sm_n = a_n\: (-1)^{\sum_{m<n}\,\ap_m\,a_m}\;,\hspace{1cm}
\sp_n = (-1)^{\sum_{m<n}\,\ap_m\,a_m}\:\,\ap_n\, .
\label{eqnarray:eqJW}
\ee
The Hamiltonian becomes:
\ba
H &=& - \frac{\e}{2}\,\sum_{n=1}^{L-1}\Biggl[
        \biggl(\e-\frac{1}{\e}\biggr)
        \biggl(\ap_n \ap_{n+1}\,-\,a_n a_{n+1}\biggr)
       +\biggl(\e+\frac{1}{\e}\biggr)
        \biggl(\ap_n a_{n+1}\,-\,a_n \ap_{n+1}\biggr)\nn\\
& &  +4\,\ap_n a_n \,-2\,-\e-\frac{1}{\e}\Biggr]\nn\\
& & +\frac{\e}{2}\,(-1)^N \,\Biggl[
     \biggl(\e-\frac{1}{\e}\biggr)
     \biggl(\ap_L \ap_{1}\,-\,a_L a_{1}\biggr)
    +\biggl(\e+\frac{1}{\e}\biggr)
      \biggl(\ap_L a_{1}\,-\,a_L \ap_{1}\biggr)\nn\\
 & & +4\,\ap_L a_L \,-2\,-\e-\frac{1}{\e}\Biggr]
\label{eqnarray:HamN}
\ea
where $N$ denotes the number of excitations:
\be
N\:=\:\sum_{m=1}^L\,\ap_m\,a_m\;.
\ee
Since the operator of total charge:
\be
Q=(-1)^{N}
\ee
commutes with $H$,
the operator can be replaced by its eigenvalue.
Therefore we separately examine the two sectors which are characterized by
the values of $Q$ \cite{Lieb}:
for $Q=-1$ we have to take periodic boundary conditions in the fermions:
\be
a^{\pm}_{L+1}\;=\;a_1^{\pm}\;,
\ee
for $Q=+1$, we have to take antiperiodic boundary conditions:
\be
a^{\pm}_{L+1}\;=\;-a_1^{\pm}\;.
\ee
After performing a Fourier transformation on the  $a_n$ and $a_n^+$:
\be
\ap_n = L^{-\frac{1}{2}}\:e^\frac{i\pi}{4}\:\sum_q\: e^{-iqn}\:\cp_q\;,
\hspace{1cm}
a_n  = L^{-\frac{1}{2}}\:e^{-\frac{i\pi}{4}}\:\sum_q\: e^{iqn}\:c_q
\label{eqnarray:Hd}
\ee
the Hamiltonian (\ref{eqnarray:HamN}) becomes:
\ba
H & =&\: -\frac{\e}{2}\:\sum_{q}\:\biggl[2\:
                            \biggl(\e-\frac{1}{\e}\biggr)\:\sin(q)\,
                                 (\cp_q\cp_{-q}+\,c_q\,c_{-q})\nn\\
         &   & +\biggl(2\:\biggl(\e+\frac{1}{\e}\biggr)\:\cos(q)-4\biggr)\,
                                  (\cp_q\,c_q+\,\cp_{-q}\,c_{-q})
                               +4-2\,\e-\frac{2}{\e}\biggr]\;,
\label{eqnarray:eqHam}
\end{eqnarray}
where $q$ takes different values depending on the total charge $Q$ and the
lattice length $L$:
\be
q=\left\{\begin{array}{cll}
            \mbox{$\frac{(2k+1)\,\pi}{L}$} & k=0,1,2,\cdots,\frac{L-2}{2}&
\mbox{$Q=+1$, L even}\\[3mm]
             \frac{(2k+1)\,\pi}{L} & k=0,1,2,\cdots,\frac{L-1}{2}&
\mbox{$Q=+1$,
L odd}\\[3mm]
             \frac{2k\,\pi}{L} & k=0,1,2,\cdots,\frac{L}{2}
& \mbox{$Q=-1$,  L even}\\[3mm]
             \frac{2k\,\pi}{L} & k=0,1,2,\cdots,\frac{L-1}{2}& \mbox{$Q=-1$,
L odd}
\end{array}
\right\} \;.
\label{eq:q}
\ee
Looking at Eq. (\ref{eqnarray:eqHam}), we notice that $H$ can be written as
$H\:=\: \sum_{0\leq q \leq \pi }\,H_q$ and has block-diagonal structure.
Each block represents a two-fermion system (the fermions having
momentum $q$ and $-q$) and
is generated by acting with $\cp_q$ and $\cp_{-q}$ on the ground state
$\vert\mbox{vac}\rangle$. This ground state is defined to be annihilated
by $c_q$:
\be
c_q\:\vert \mbox{vac}\rangle\:=\,0\; ;
\ee
so
$\vert\mbox{vac}\rangle$ corresponds to a state without fermions.
Each block $H_q$ for $q\neq 0,\pi$ can be written as a $4\times 4$ matrix.
Choosing the
basis $\{\vert\mbox{vac}\rangle, \cp_q
\:\vert\mbox{vac}\rangle,
\cp_{-q}\:\vert\mbox{vac}\rangle,
\cp_q\,\cp_{-q}\:\vert\mbox{vac}\rangle\}$, it takes the form:
\[ \left( \begin{array}{clll}
0  & & & \e(\e-\frac{1}{\e})\sin{q}\\
   & -\e((\e+\frac{1}{\e})\cos{q}-2) & &\\
   & &  -\e((\e+\frac{1}{\e})\cos{q}-2) & \\
\e(\e-\frac{1}{\e})\sin{q} & & & -2\e((\e+\frac{1}{\e})\cos{q}-2)
\end{array} \right)\;. \]
$H_{0}$ and $H_{\pi}$ are
$2 \times 2$ diagonal matrices: in the basis $\{\vert\mbox{vac}\rangle, \cp_{0}
\:\vert\mbox{vac}\rangle   \}$:
\[ H_{0}=\left( \begin{array}{cl}
0 & \\
& -\e\,(\e+\frac{1}{\e})+2\e
\end{array} \right) \]
and
in the basis $\{\vert\mbox{vac}\rangle, \cp_{\pi}
\:\vert\mbox{vac}\rangle   \}$:
\[ H_{\pi}=\left( \begin{array}{cl}
0 & \\
& \e\,(\e+\frac{1}{\e})+2\e
\end{array} \right)\;. \]
The eigenvalues of the $4 \times 4$ matrices $H_q$ are:
\ba
\l_{1,2} & = & -\e\le[\le(\e+\frac{1}{\e}\ri)\cos(q)-2\ri]\;,\\
\l_{3}   & = & -\e\le[\le(\e+\frac{1}{\e}+2\ri)\cos(q)-
                      \le(\e+\frac{1}{\e}+2\ri)\ri]\;,\\
\l_{4}   & = & -\e\le[\le(\e+\frac{1}{\e}-2\ri)\cos(q)
                     -\le(\e+\frac{1}{\e}-2\ri)\ri]\;.
\ea
They can be expressed in terms of free fermion energies by adding a shift of:
\be
s(q)= \e\le[\le(\e+\frac{1}{\e}-2\ri)\cos(q)+\le(\e+\frac{1}{\e}-2\ri)\ri]
\ee
to each eigenvalue, resulting in:
\ba
\l_{1,2} & = & \e\le(\e+\frac{1}{\e}-2\cos(q)\ri)\;,\\
\l_{3}   & = & 2 \e\le(\e+\frac{1}{\e}-2\cos(q)\ri)\;,\\
\l_{4}   & = & 0\;.
\ea
Thus in each block $\l_{1,2}$ can be identified as one-fermionic
excitations, $\l_{3}$
corresponds to a two-fermionic excitation and $\l_{4}$ to the state without
fermions.

The eigenvalues of $H_{0}$ have to be shifted by an amount of $\frac{s(0)}{2}$
so that we obtain:
\ba
\l_1(q=0) & =& 0\;,\nn\\
\l_2(q=0) & =& \e\le(\e+\frac{1}{\e}-2\ri)\;.
\ea
Again $\l_2(q=0)$ represents a one-fermionic excitation with momentum $0$.
The eigenvalues of $H_{\pi}$ already have the desired form of
fermionic excitations.

In this way one obtains a set of free fermions, each fermion having an energy
of:
\be
\l_q^F \:=\: \e\le(\e+\frac{1}{\e}-2\cos(q)\ri)\;,
\label{eq:eq.FE}
\end{equation}
where $q$ takes not only the values (\ref{eq:q}) but also the corresponding
negative values except for $q=0,\pi$ where only the positive value of
$q$ appears.

Let now $\c^{\pm}_q$ be the corresponding set of fermionic operators,
obeying the
anticommutation relation:
\be
\{\c^{\mu}_q,\c^{\nu}_k\}=\d_{\mu,-\nu}\: \d_{q,k}
\ee
so that $H$ can be written in the diagonal form:
\be
H = P^{+}\sum_{q_{\mbox{\scriptsize even}}}\,\l^F_q\,\c^+_q\c_q\,P^{+}\,
   +\,P^{-}\sum_{q_{\mbox{\scriptsize odd}}}
  \l^F_{q}\,\c^+_{q}\,\c_{q}\,P^{-}\;.
\ee
In this expression, $P^{\pm}$ denote the projection operators onto the sectors
$Q=\pm1$ respectively, $q_{\mbox{\scriptsize even}}$ and
$q_{\mbox{\scriptsize odd}}$ the values $q$ takes in these sectors.

{\sl All excitations are only even combinations of fermions},
in the sector with $Q=+1$ as well as in the sector with $Q=-1$.
This can be understood taking into account that the ground
state (with energy zero) in the sector with $Q=-1$ has already charge $-1$.
So all the excitations in this sector have to be even combinations of the
fermion energies (\ref{eq:eq.FE}). The same argument holds for the $Q=+1$
sector, where the ground state has charge $+1$.
%-----------------------------------------------------
%
\section{Transformations between Two-State Models}
\label {sec:connex}
\hspace{\parindent}For
a given set of rates in the coagulation-decoagulation model one might
think that it is possible to choose the rates in the annihilation-creation
model in such a way that the Hamiltonians (\ref{eq:HCD3}) and (\ref{eq:HAC3})
become identical up to an overall factor.
In general this leads to negative and thus unphysical rates. But if
the backward reactions decoagulation and creation do not occur,
we can get a mapping between Hamiltonians describing reactive-diffusive
systems as will be shown now.
We consider:
\begin{itemize}
\item The coagulation model with:
\ba
\mbox{coagulation-rate } \wio(0,1) & = & \mbox{arbitrary},\nn\\
\mbox{decoagulation-rate } \wio(1,1) & = & 0\;. \nn
\ea
Without any transformation, the Hamiltonian (\ref{eq:HCD1})
splits into two parts $H_{\mbox{\scriptsize coag.}}=H_0+H_1$, where
the spectrum is determined by $H_0$ alone:
\ba
H_0 & = & -\half\sum_{i=1}^{L}\Bigl[
        \sx_i\sx_{i+1}+\sy_i\sy_{i+1}+\D'\sz_i\sz_{i+1}
      +(1-\D')(\sz_i+\sz_{i+1})+\D'-2\Bigr]
\label{eq:HDC=0}\\
H_1 & = & -\half(1-\D')\sum_{i=1}^{L}
              \Bigl[\sp_i(1-\sz_{i+1})+(1-\sz_i)\sp_{i+1}\Bigr]
\ea
with:
\be
\D'  =  \D = 1-\wio(0,1)\;.
\label{eq:DelC}
\ee
\item The annihilation model with:
\ba
\mbox{annihilation-rate } \wii(0,0) & = & \mbox{arbitrary},\nn\\
\mbox{creation-rate } \wii(1,1) & = & 0\;. \nn
\ea
The Hamiltonian (\ref{eq:HAC1}) then becomes
$H_{\mbox{\scriptsize ann.}}=H_0+H_1$ (again without
any transformation, again the spectrum given by $H_0$ alone):
\ba
H_0 & = & -\half\sum_{i=1}^{L}\Bigl[
        \sx_i\sx_{i+1}+\sy_i\sy_{i+1}+\D'\sz_i\sz_{i+1}
      +(1-\D')(\sz_i+\sz_{i+1})+\D'-2\Bigr]
\label{eq:HCR=0}\\
H_1 & = & -(2-2\D')\sum_{i=1}^{L}\sp_i\sp_{i+1}
\label{eq:H1CR0}
\ea
where:
\be
\D' = \D = 1-\frac{\wii(0,0)}{2}\;.
\label{eq:DelA}
\ee
\end{itemize}
Inspecting the Eqs. (\ref{eq:DelC}) and (\ref{eq:DelA}), it is obvious
that the two models have identical spectra,
if the rates are chosen in a way that:
\be
\frac{\wii(0,0)}{2} = \wio(0,1)\;,
\label{eq:equicond}
\ee
i.e.: the annihilation-rate has to be twice the coagulation-rate.
If we want to obtain the spectrum of these Hamiltonians from the fermionisation
presented in Section \ref{sec:diag}, we have to choose the rates in a way
that:
\be
\D' = 0\;,
\ee
i.e.: coagulation-rate=1, annihilation-rate=2.

We now give the transformation between the two Hamiltonians explicitly.
Because $H_{\mbox{\scriptsize ann.}}$ is $Z_2$-symmetric (i.e. commutes with
the operator $\prod_{i=1,\ldots,L}\sz_i$) there are two possibilities.
With:
\be
b_1 = \le( \begin{array}{cc}
         1 & -1\\
         0 &  2
    \end{array} \ri)\;,\hspace{1cm}
b^{-1}_1 = \le( \begin{array}{cc}
         1 & \half\\
         0 & \half
    \end{array} \ri)\
\label{eq:b1}
\ee
or
\be
b_2 = \le( \begin{array}{cc}
         1 &  1\\
         0 & -2
    \end{array} \ri)\;,\hspace{1cm}
b^{-1}_2 = \le( \begin{array}{cc}
         1 &  \half\\
         0 & -\half
    \end{array} \ri)
\ee
we define the $L$-fold tensor product:
\be
B = b^{\ten L}\;,\hspace{1cm} B^{-1}=(b^{-1})^{\ten L}\;.
\ee
Then we have (cf. Appendix \ref{sec:simpro} for the proof):
\be
H_{\mbox{\scriptsize coag.}} = BH_{\mbox{\scriptsize ann.}}B^{-1}\;.
\ee
Note that $b_1\sz=b_2$.
The transformation between the Hamiltonians of the annihilation model and
the coagulation model is non-unitary but still local.
The existence of the similarity transformation proves
the conjecture that both models are equivalent \cite{Kang}.

As we will see now, $b_1$ is the correct choice
to obtain a transformation that maps physically acceptable initial
states of one model onto acceptable initial states of the
other.
\subsection{Transformation of Expectation Values}
\label{sec:Traexval}
\hspace{\parindent}To
study in which way expectation values are changed by the
transformation $B=b^{\ten L}_1$, we first restrict ourselves to initial
probability distributions which have product form:
\be
P_0(\{\b\}) = \prod_{i=1}^L\le[p_i\d(\b_i-1)+(1-p_i)\d(\b_i)\ri]\;.
\label{eq:prod_state}
\ee
If we take $0\leq p_i\leq 1$, this is a state where site
$i$ is occupied with probability $p_i$.
To simplify the notation, we define the vector of initial occupation
probabilities:
\be
\vec{P}_0 = (p_1,\ldots,p_L)\;.
\label{eq:pstate}
\ee
The corresponding initial ket state then is given by:
\be
|\vec{P}_0\rangle = {{1-p_1}\choose{p_1}}\ten\cdots\ten{{1-p_L}\choose{p_L}}\;.
\label{eq:pvecstat}
\ee
The expectation value of an operator $X$
is a function of time and depends on the initial condition.
We use the notation $<X>(\vec{P}_0,t)$ to remind of these dependencies.
The transformation behaviour can now be calculated:
\ba
<X>_{\mbox{\scriptsize coag.}}(\vec{P}_0,t) & = &
 \langle 0|X\,e^{-H_{\mbox{\scriptsize coag.}} t}|\vec{P}_0\rangle\\
& = &
 \langle 0|XB\,e^{-H_{\mbox{\scriptsize ann.}} t}B^{-1}|\vec{P}_0\rangle
\label{eq:xb2}\\
& = &
\Bigl\langle 0\Bigr|XB\,e^{-H_{\mbox{\scriptsize ann.}} t}\Bigl|
\half\vec{P}_0\Bigr\rangle\label{eq:xb3}\\
& = & <XB>_{\mbox{\scriptsize ann.}}\le(\half\vec{P}_0,t\ri)\;.
\label{eq:XBtraf}
\ea
To derive Eq. (\ref{eq:xb3}) from Eq. (\ref{eq:xb2}) one uses the explicit
representations (\ref{eq:b1}) for $B^{-1}$ and (\ref{eq:pvecstat}) for
$|\vec{P}_0\rangle$. The result can be summarized as follows:
in order to obtain the expectation value of an operator $X$ in
the coagulation model,
one can calculate the expectation value of $XB$ in the annihilation model
with the initial occupation probabilities devided by two.

Let now $X$ be a product of occupation number operators:
\be
n_i = \le( \begin{array}{cc}
         0 &  0\\
         0 &  1
    \end{array} \ri)_i\;.
\ee
Because $n_i\,b = 2n_i$ and $\overline{b} = 1$ (cf. Eq.
(\ref{eq:exval2}) for the relation of $b$ and $\overline{b}$) we derive
from Eq. (\ref{eq:XBtraf}):
\ba
<n_{i_1}\cdots n_{i_k}>_{\mbox{\scriptsize coag.}}(\vec{P}_0,t) & = &
<(n_{i_1}\cdots n_{i_k})B>_{\mbox{\scriptsize ann.}}
            \le(\half\vec{P}_0,t\ri)\nn\\
 & = & <\overline{(n_{i_1}\cdots n_{i_k})B}>_{\mbox{\scriptsize ann.}}
            \le(\half\vec{P}_0,t\ri)\nn\\
 & = & 2^k<n_{i_1}\cdots n_{i_k}>_{\mbox{\scriptsize ann.}}
            \le(\half\vec{P}_0,t\ri)\;.
\label{eq:Corrtra}
\ea
As a special case, we obtain for the concentration
$c=\frac{1}{L}\sum_{i=1}^L n_i\;$:
\be
c_{\mbox{\scriptsize coag.}}(\vec{P}_0,t) =
2\, c_{\mbox{\scriptsize ann.}}\le(\half\vec{P}_0,t\ri) \;.
\label{eq:konztra}
\ee
For homogeneous initial occupation probability $p_1=p_2=\cdots=p_L=p$ this
result was already obtained by \cite{Privman}.

The generalization to arbitrary initial probability distribution is
straightforward, because every initial state is a linear combination
of product states:
\ba
P_0(\{\b\}) & = & \sum_{\{\b'\}}P_0(\{\b'\})
                  \prod_{i=1}^L\le[\b'_i\d(1-\b_i)+(1-\b'_i)\d(\b_i)\ri]\;.
\ea
The values the $p_i$ take in each term are thus only 1 or 0.
This formula simply reflects the fact that the basis vectors
$|\{\b\}\rangle$ can be written in the form (\ref{eq:pvecstat}) with
$p_i=\b_i$. The application of the similarity transformation then yields:
\be
(B^{-1}P_0)(\{\b\})=\sum_{\{\b'\}}P_0(\{\b'\})
                  \prod_{i=1}^L\le[\frac{\b'_i}{2}\d(1-\b_i)
                                  +\le(1-\frac{\b'_i}{2}\ri)\d(\b_i)\ri]\;.
\ee
As long as the interpretation of $|\vec{P}_0\rangle$ as a probability
distribution never enters the calculations of expectation values, one
can confidently use all results obtained in the coagulation model for
the annihilation model. This is very surprising, because one has
to use unphysical ``probabilities'' between 1 and 2 in the coagulation
model if the probabilities in the annihilation model take values between
1/2 and 1. Because the transformation $B$ is local, all results obtained
in this section are valid for periodic as well as for open boundary conditions.
%
%--------------------------------------------------------
%
%
\section{Probabilistic Approach to the \protect \\
         Coagulation-Decoagulation Model}
\label{sec:holes}
\hspace{\parindent}As
we observed in Section \ref{sec:diag}, the spectrum of the Hamiltonian
describing a sytem with coagulation rate = diffusion rate = 1 and an
arbitrary decoagulation rate is known. But nevertheless, the
complicated structure of the
eigenvectors and the involved matrix elements makes it extremly
difficult to calculate
observables like the mean value of the particle number directly from the
time evolution of the probability distribution $P(\{\b\},t)$.
Following \cite{ben-Avraham2} we use a different approach where one does
not have to deal with a space of $2^L$ eigenvectors but only with
$L$ subspaces of $(L-1)$ degrees of freedom.
Quantities of interest,
such as the expectation value of the occupation number for each site, can
already be calculated from these subspaces.
\subsection{Empty Interval Probabilities}
\hspace{\parindent}We
introduce the empty interval probability $\O(j,n,t)$, which is
the probability to find the $n$ consecutive sites
$j-\frac{n}{2}+1,\ldots,j+\frac{n}{2}$
empty at time $t$. Because we have chosen periodic boundary conditions,
the arithmetic operations on the numbering of the sites are performed
modulo $L$.
Notice that $j$ takes half integer values if $n$ is
odd. For example $\O(5/2, 3, t)$ is the probability to find
sites 2, 3 and 4 empty at time $t$.
The probability to have site $j$ occupied is clearly: $1-\O(j-\half,1,t)$.
Therefore we can calculate the concentration as:
\be
c(t) = 1- \frac{1}{L}\sum_{j=\half}^{L-\half}\O(j,1,t)\;.
\ee
Using the quantum chain picture introduced in Section \ref{sec:master},
the time evolution of the system is given by a Hamiltonian
$H=\sum H_i$. $H_i$ operates locally in the tensor product of the
two-dimensional vector spaces attached to site $i$ and $i+1$.
The matrix representation of $H_i$ is given by:
\be
H_i=\le(\begin{array}{cccc} 0   &   0   &   0   &   0\\
                            0   & \e^2  &  -1   &  -1\\
                            0   &  -1   & \e^2  &  -1\\
                            0   &1-\e^2 &1-\e^2 &   2
    \end{array}\ri)\;.
\ee
Here we used the notation $\e^2=1+\woi(1,1)$.
$\O(j,n,t)$ is calculated as the expectation value of the
empty interval operator $O(j,n)$:
\be
<O(j,n)>(t) = \O(j,n,t)\;
\ee
where
\be
O(j,n) = \le(\begin{array}{cc}
             1 & 0 \\
             0 & 0 \end{array}\ri)_{j-\frac{n}{2}+1}
         \ten\cdots\ten
         \le(\begin{array}{cc}
             1 & 0 \\
             0 & 0 \end{array}\ri)_{j+\frac{n}{2}}\;.
\ee
Now we can calculate an equation describing the time evolution of $\O(j,n,t)$.
We first consider the cases $n=2,\ldots,L-1$:
\ba
\frac{\pa}{\pa t}\O(j,n,t)
& = & \frac{\pa}{\pa t} \langle 0|O(j,n)|P(t)\rangle\\[1mm]
& = & \frac{\pa}{\pa t} \langle 0|O(j,n)\,e^{-H t}|P(0)\rangle\\[1mm]
& = & -\langle 0|O(j,n)\,H|P(t)\rangle\\[1mm]
& = & -\sum_{i=1}^L\langle 0|O(j,n)\,H_i|P(t)\rangle
\label{eq:h1}\\[1mm]
& = & -\langle 0|O(j,n)\,H_{j-n/2}|P(t)\rangle
      -\langle 0|O(j,n)\,H_{j+n/2}|P(t)\rangle
\label{eq:h2}\\[1mm]
& = &  \e^2\langle 0|O(j-1/2,n+1)|P(t)\rangle
      +\e^2\langle 0|O(j+1/2,n+1)|P(t)\rangle\nn\\[1mm]
&   & -2(\e^2+1)\langle 0|O(j,n)|P(t)\rangle\nn\\[1mm]
&   & +\langle 0|O(j-1/2,n-1)|P(t)\rangle
      +\langle 0|O(j+1/2,n-1)|P(t)\rangle\;.
\label{eq:h3}
\ea
To derive Eq. (\ref{eq:h2}) from Eq. (\ref{eq:h1}) we simplified the
expectation values in the way introduced in Sec.~\ref{sec:master},
Eq. (\ref{eq:exval2}).
For $i=1,\ldots,j-\frac{n}{2}-1$ and $i=j+\frac{n}{2}+1,\ldots,L$
$H_i$ operates directly
to the left on the bra ground state $\langle 0|$. Therefore we can use
the identity:
\be
\overline{H}_i = 0
\ee
and these terms vanish.
For $i=j-\frac{n}{2}+1,\ldots,j+\frac{n}{2}-1$ we have:
\be
\le[\le(\begin{array}{cc}
             1 & 0 \\
             0 & 0 \end{array}\ri)_{i}
         \ten
         \le(\begin{array}{cc}
             1 & 0 \\
             0 & 0 \end{array}\ri)_{i+1}\ri]\,H_i = 0\;.
\ee
Therefore there are only two terms left, and Eq. (\ref{eq:h2}) follows.
Applying Eq. (\ref{eq:exval2}) several times, we finally get Eq. (\ref{eq:h3}).
For $n=1$ we get:
\ba
\frac{\pa}{\pa t}\O(j,1,t)
& = &  \e^2\langle 0|O(j-1/2,2)|P(t)\rangle
      +\e^2\langle 0|O(j+1/2,2)|P(t)\rangle\nn\\[1mm]
&  &  -2(\e^2+1)\langle 0|O(j,1)|P(t)\rangle\nn\\[1mm]
&  &  +\langle 0|P(t)\rangle
      +\langle 0|P(t)\rangle\;.
\label{eq:h4}
\ea
Because $H$ conserves the sum of probabilities, we have:
\be
\langle 0|P(t)\rangle = \langle 0|P(0)\rangle = 1.
\label{eq:norm}
\ee
Notice that every state of the form (\ref{eq:pstate})
obeys  $\langle 0|\vec{P}_0\rangle=1$, irrespectively of the values
the $p_i$ take.
Therefore the case $n=1$ is included in Eq. (\ref{eq:h3}) if we formally put
$O(j,0) = 1$, i.e.:
\be
\O(j,0,t) = 1\;.
\label{eq:Onull}
\ee
In summary we get the following differential equations:
\ba
\frac{\pa}{\pa t}\O(j,n,t) & = & +\e^2\:\O\biggl(j-\half,n+1,t\biggr)\:
                                  +\:\e^2\:\O\biggl(j+\half,n+1,t\biggr)\nn\\*
                            &   & -2(\e^2+1)\:\O(j,n,t)\nn\\*
                            &   & +\O\biggl(j-\half,n-1,t\biggr)\:
                                  +\:\O\biggl(j+\half,n-1,t\biggr)\;,
\label{eq:hol1}\\[3mm]
n & = & 1,\ldots,L-1\;\;,\nn\\
j & = & \le\{
\begin{array}{ll}
      \half,\ldots,L-\half & \mbox{if n is odd}\\
      1,\ldots,L           & \mbox{if n is even}\;.
\end{array}
\ri.\nn
\ea
For $n=L$ we have in addition:
\be
\frac{\pa}{\pa t}\O(j,L,t) = 0\;.
\label{eq:holGL}
\ee
This leads to:
\be
\O(j,L,t) = \O(j,L,0) = \mbox{const}
\label{eq:OL}
\ee
for all $j$ (note that all $\O(j,L,t)$ are equal).
The general case of chemical systems, for wich a system of differential
equations for empty interval probabilities can be obtained, will be
discussed by \cite{Peschel}.
Our derivation of the differential equations differs from the one presented
in \cite{ben-Avraham2,Peschel}.
We prefered a purely algebraic way, because it shows
to what extend the results depend on the fact that one deals with
probabilities. The similarity transformation may be used to
get results for the annihilation model because the fact that unphysical
``probabilities'' (cf. Sec. \ref{sec:Traexval}) occur
does not affect the normalisation (\ref{eq:norm}).

Because of translation invariance we can
use a Fourier transformation with respect to the center-coordinate $j$:
\be
\Oh(k,n,t) = \frac{1}{L}\sum_{j}\O(j,n,t)\exp\biggl(i\frac{2\pi k}{L}j\biggr)
             \;;\hspace{1cm} k = 0,1,\ldots,L-1\;\;.
\ee
The boundary conditions (\ref{eq:Onull}) and (\ref{eq:OL}) lead to:
\ba
\Oh(k,L,t) & = & \le\{
\begin{array}{ccl}
\O(\frac{L}{2},L,0) & \mbox{if} & k = 0\\
     0                & \mbox{if} & k = 1,\ldots,L-1\;\;;
\end{array}
\label {eq:OhL}
\ri.\\[2mm]
\Oh(k,0,t) & = & \le\{
\begin{array}{lll}
     1 & \mbox{if} & k = 0\\
     0
& \mbox{if} & k = 1,\ldots,L-1\;\;.
\end{array}
\label {eq:Oh0}
\ri.
\ea
The $\Oh(k,n,t)$ satisfy the following differential equations:
\ba
\frac{\pa}{\pa t}\Oh(k,n,t)
 & = & 2\e^2\cos\biggl(\frac{\pi k}{L}\biggr)\:\Oh(k,n+1,t)\nn\\*
 &   & -2(\e^2+1)\:\Oh(k,n,t)\nn\\*
 &   & +2\cos\biggl(\frac{\pi k}{L}\biggr)\:\Oh(k,n-1,t)\;\;,
\label{eq:holFT}
\ea
\[n = 1,\ldots,L-1\;\;,\hspace{1cm}
  k = 0,\ldots,L-1\;\;.
\]
Thus for each $k$ there is a system of $L-1$ equations, which is inhomogeneous
only for $k=0$ because of $\Oh(0,0,t)$ and $\Oh(0,L,t)$.
The solution is obtained in two steps. We first solve the homogeneous problem.
Then we add a particular solution in order to obtain the complete solution.
\subsection{The Homogeneous Problem}
\hspace{\parindent}The
homogeneous problem can be considered as a simple eigenvalue problem.
Expand $\Oh(k,n,t)$ in eigenfunctions of:
\ba
 \L_l^k\Psi_l^k(n)
 & = & 2\e^2\cos\biggl(\frac{\pi k}{L}\biggr)\:\Psi_l^k(n+1)\,
      -2(\e^2+1)\:\Psi_l^k(n)\,
      +2\cos\biggl(\frac{\pi k}{L}\biggr)\:\Psi_l^k(n-1)\;,\\
\Psi_l^k(0) & = & 0\;,\\
\Psi_l^k(L) & = & 0\;.
\ea
Then $\Oh(k,n,t)$ can be expressed as a linear combination of the
eigenfunctions as:
\be
\Oh(k,n,t) = \sum_l A^k_l\Psi_l^k(n)\:e^{\L_l^kt}\;.
\ee
The eigenfunctions and eigenvalues are easily derived for each $k$:
\ba
\Psi_l^k(n) & = & \e^{-n}\:\sin\biggl(\frac{\pi l n}{L}\biggr)\;,
\label{eq:holefunc}\\
\L_l^k      & = & \e\biggl[4\cos\biggl(\frac{\pi k}{L}\biggr)\:
                             \cos\biggl(\frac{\pi l}{L}\biggr)\:
                  -2(\e+\e^{-1})\biggr]
\label{eq:holeval}
\ea
where the index $l$ counts the different eigenvalues for each $k$:
\[
k = 0,\ldots,L-1\;,\hspace{1cm}
l = 1,\ldots,L-1\;.
\]
The eigenfunctions $\Psi^k_l$ obey the orthogonality relation:
\be
\frac{2}{L}\:\sum_{n=1}^{L-1}\e^{2n}\:\Psi_l^k(n)\:\Psi_{l'}^k(n) =
          \d_{l,l'}\;.
\label{eq:holot}
\ee
At this stage it is interesting to investigate the relation between
the results of the probabilistic approach and the spectrum of the
quantum chain Hamiltonian obtained in free fermions in Sec. \ref{sec:diag}:
We can express the eigenvalues (\ref{eq:holeval}) as:
\be
\L_l^k  =  \e\le[2\cos\le(\frac{\pi(k+l)}{L}\ri)\:
                +2\cos\le(\frac{\pi(k-l)}{L}\ri)
                -2\le(\e+\frac{1}{\e}\ri)\ri]\;.
\ee
Compared with the energies (\ref{eq:eq.FE}) of the fermions, these eigenvalues
can clearly be identified as the two fermionic excitations of the
quantum chain.
\subsection{The Inhomogeneous Problem}
\hspace{\parindent}Due
to the fact that $\Oh(0,L,t)$ and $\Oh(0,0,t)$ are constant, we have
to solve an inhomogenous problem in the $k=0$ sector.
We find two steady states:
\ba
\Psi^0_{\mbox{\scriptsize vac}}(n) & = & 1\;,\\
\Psi^0_{\mbox{\scriptsize z.m.}}(n) & =   & \e^{-2n}\;.
\ea
The first steady state simply corresponds to an empty lattice, i.e. to the
fermionic vacuum of the quantum chain. The second one is a non-trivial
zero mode, which
is related to the charged ground state of the negative $Q$ sector.
The following combination satisfies the boundary conditions (\ref{eq:OhL}),
(\ref{eq:Oh0}):
\be
\Psi^0_s(n)  =  \frac{1}{1-\e^{-2L}}\biggl(
                  (1-\Oh(0,L,0))\:\e^{-2n}\:+\:\Oh(0,L,0)-\e^{-2L}\biggr)\;.
\label{eq:holihs}
\ee
The complete solution is:
\be
\Oh(k,n,t)  =  \sum_{l=1}^{L-1}\Bigr[A_l^k\:\Psi_l^k(n)
                 \:\exp(\L_l^k\:t)\Bigr]\:+\:\d_{k,0}\:\Psi^0_s(n)
\label{eq:holCS}
\ee
where the coefficients $A_l^k$ are computed using the
orthogonality relation (\ref{eq:holot}):
\be
A_l^k     = \frac{2}{L}\sum_{n=1}^{L-1}\e^{2n}\:
                 \Bigl(\Oh(k,n,0)\:-\:\d_{k,0}\:\Psi^0_s(n)\Bigr)\:
                 \Psi^k_l(n)\;.
\label{eq:holFC}
\ee
This allows to calculate the time evolution of the empty
interval probabilities for every lattice length $L$ and arbitrary
initial conditions. We proceed with a special choice for the initial condition.
\subsection{The Concentration for Uncorrelated Initial Conditions}
\label{sec:init}
\hspace{\parindent}Using
the fact that $1-\O(j-\half,1,t)$ is the probability to have site $j$
occupied at time $t$ the mean value of the particle concentration is
found to be:
\ba
c\:(t) & = & \frac{1}{L}\:\sum_{j=\half}^{L-\half}[1-\O(j,1,t)]\nn\\
     & = & 1-\Oh(0,1,t)\;.
\ea
The coefficients $A^k_l$ (Eq. (\ref{eq:holFC})) can be calculated if we use
a product state of the form (\ref{eq:prod_state}) as initial state.
Let $p$, the probability to have a site
occupied at time $t=0$, be the same for all sites. Then we have:
\be
\Oh(0,n,0) = \O(j,n,0) = (1-p)^n\;
\ee
and
\be
A_l^0  =  \frac{2}{L}\biggl[\frac{1+(-1)^{l+1}\:\e^L\:(1-p)^L}
             {\frac{\e ^{2}(1-p)^2+1}{\e (1-p)}-2\cos(\frac{\pi l}{L})}\nn\\
           -\frac{1+(-1)^{l+1}\:\e^L\:(1-p)^L}
              {\frac{\e ^{2}+1}{\e}
                -2\cos(\frac{\pi l}{L})}\biggr]
                 \sin\biggl(\frac{\pi l}{L}\biggr)\;.
\ee
The mean value of the concentration is given by:
\be
c\:(t,L)  =  \frac{(1-\e^{-2})\:(1-(1-p)^L)}{1-\e^{-2L}}
           -\sum_{l=1}^{L-1}\;A_l^0\:\e^{-1}
            \sin\biggl(\frac{\pi l}{L}\biggr)\:
            \exp(\L^0_l\:t\:)
\label{eq:holc(t)}
\ee
where the $\L^0_l$ are calculated from Eq. (\ref{eq:holeval}):
\be
\L_l^0
  =  \e\biggl[4\:\cos\biggl(\frac{\pi l}{L}\biggr)\:-2(\e+\e^{-1})
               \biggr]
\;.
\ee
This expression for the concentration is exact for all lattice sizes $L$.
%
%-------------------------------------------------------
%
% Kapitel ueber Finite-Size-scaling
\section {Finite-Size Scaling of the Coagulation Model}
\label{sec:fin_size}
\hspace{\parindent}In
this section we will
study the finite-size scaling behaviour of the coagulation model.
This means taking the
limit $L\rightarrow\infty$, $t\rightarrow\infty$ while the scaling variable
$z\,=\,\frac{4t}{L^2}$ is kept fixed.
As has been explained
in the introduction, the expansion of the concentration in the scaling limit
Eq. (\ref{eq:scal1}) is:
\be
c(z,L)\;=\;L^x\;[F_{0}(z)+L^{-y}F(z)+\cdots]
\label{eq:scal2}
\ee
where $F_{0}$ is the scaling function and $x$ the scaling exponent.
We will determine the correction term $L^{-y}F$ as well.
In the limit $\e=1$ (no decoagulation) we derive from Eq.
(\ref {eq:holc(t)}):
\be
c(t,L) \;=\; \frac{1-(1-p)^{L}}{L}\,-\,\sum_{l=1}^{L-1}\,A_l^{0}\,
               \sin \le(\frac{\pi l}{L}\ri) \exp (\l_l^{0} t)
\label{eq:c}
\ee
with
\ba
A^{0}_{l} & = & \frac{1}{L}\:\sin\biggl(\frac{\pi\,l}{L}\biggr)\:
                \Biggl[\frac{1+(-1)^{l+1}(1-p)^{L}}{\frac{1+(1-p)^{2}}
                            {2\,(1-p)}-\cos(\frac{\pi\,l}{L})}

-\frac{1+(-1)^{l+1}(1-p)^{L}}{1-\cos(\frac{\pi\,l}{L})}\Biggr]\\
\l^{0}_{l}& = & 4\:\biggl(\cos\biggl(\frac{\pi\,l}{L}\biggr)-1\biggr).
\ea
For large $L$ the term $(1-p)^{L}$ is neglegiable
so that Eq. (\ref{eq:c}) becomes:
\ba
c(z,L) & = & \frac{1}{L}\:+\frac{1}{L}\:\sum_{k=1}^{L-1}\;
            \biggl(1+\cos\:\biggl(\frac{\pi k}{L}\biggr)\biggr)\:
           \exp\:\biggl[\:-z\,L^2\:\biggl(1-\cos\:\biggl(\frac{\pi k}{L}\biggr)
             \biggr)\biggr]\nn\\
    &    & -\frac{1}{L}\:\sum_{k=1}^{L-1}\:\frac{\sin^2\: (\frac{\pi k}{L})}
               {\:\frac{1+(1-p)^2}{2\:(1-p)}-\cos\:(\frac{\pi k}{L})}\:
            \exp\:\biggl[\:-z\,L^2\:\biggl(1-\cos\:\biggl(\frac{\pi k}{L}
                       \biggr)\biggr)\:\biggr]\;.
\label{eq:conz}
\ea
Now performing the scaling limit, it is possible to write $L\:c(z,L)$ as a
power series in $\frac{1}{L}$ with coefficients depending only on the scaling
variable $z$:
\be
L\:c(z,L) = \:F_{0}\:(z)+\frac{1}{L^2}\:F\:(z)+O\:\biggl(\frac{1}{L^4}
             \biggr)\;.
\label {eq:eq1.3}
\ee
The coefficients can be identified as Jacobi theta functions and their
derivatives:
\ba
L\:c(z,L) & = & \th_3\:\biggl(0,\frac{i\pi z}{2}\biggr)\nn\\
        &   & +\frac{1}{L^2}\:\biggl[\:\frac{z}{6}
                 \:\frac{\partial^2}{\partial z^2}\:
                  \th_3\:\biggl(0,\frac{i\pi z}{2}\biggr)
                  +\frac {(p-2)^2}{2\:p^2}\:\frac{\partial}{\partial z}\:
                  \th_3\:\biggl(0,\frac{i\pi z}{2}\biggr)\biggr]\:
              +O\:\biggl(\frac{1}{L^4}\biggr)
\label{eq:scf}
\ea
where $\th_3(u,\t)$ is defined by the series:
\be
\th_3\:(u,\t)\:=\:\sum_{l=-\infty}^{\infty}\:e^{i\pi\t l^2}\:e^{2i l u}\;.
\label {eqn:eq.deft}
\ee
Concerning the initial assumption Eq. (\ref{eq:scal1}) or Eq. (\ref{eq:scal2})
about the form of the
concentration in the finite-size limit, the critical scaling exponent $x$ is
found to be equal to $-1$, the correction exponent $y=2$.
Observe that the initial probability $p$ (to find a particle at a site at time
$t=0$) enters only in the corrections. So the scaling function for
this model is independent of the initial
conditions as far as an uncorrelated initial state is concerned.
It reflects the phenomenon of self-organisation:
with increasing time and lattice length, the influence of the
initial conditions vanishes.

At this point, it is interesting to mention that it is possible as well
to obtain the
scaling function, using the continuum
limit of the lattice model following \cite{ben-Avraham2}.
These calculations will be presented in Appendix \ref{sec:Cont}.

Before closing this section, we relate the concentration
in the scaling limit, taking additionally small $z$, to
the long-time behaviour of the concentration in the thermodynamic limit.
For this purpose, we apply the Poisson resummation formula to the scaling
function $\th_3(0,\frac{i \pi z}{2})$:
\be
\th_3\le(0,\frac{i \pi z}{2}\ri)\; = \; \sqrt{\frac{2}{\pi z}}\;
\biggl[\,1\,+\,2\;\sum_{m=1}^{\infty}\,
\exp\le(-\frac{2m^2}{z}\ri) \biggr]
\label{eq:poi}
\ee
With the help of this expression, we can now take the limit for small $z$
in Eq. (\ref{eq:scf}):
\be
c(t) = \frac{1}{L}\sqrt{\frac{2}{\pi z}}\le[ 1
      -\frac{1}{4zL^2}\le(\frac{(p-2)^2}{p^2}-\frac{1}{2}\ri)+\cdots\ri]\;.
\ee
Inserting the definition of $z$ yields:
\be
c(t) = \sqrt{\frac{1}{2\pi t}}\le[ 1
      -\frac{1}{16t}\le(\frac{(p-2)^2}{p^2}-\frac{1}{2}\ri)+\cdots\ri]\;.
\label{eq:ltime}
\ee
The leading term was already obtained by \cite{ben-Avraham2,Lin}.

This relationship illustrates the fact already mentioned by
\cite{Alcaraz}: the scaling exponent $x=-1$ is
twice the exponent $\alpha$ of $t$ in the long-time behaviour of the
concentration in the thermodynamic limit  where
$c(t) \simeq t^{\alpha}$.

This scaling relation can be understood from a more general point of view
looking at the scaling behaviour
of the concentration. In the thermodynamic limit $L\rightarrow \infty$
while $t$ is kept fixed, $z$ becomes small.
The scaling function and the corrections can be developed in powers
of $\frac{1}{z}$ (cf. Eq.(\ref{eq:poi})):
\ba
c(z,L)&=& L^x \;F_{0}(z)+\ldots\\
      &=& L^x \le(\frac{1}{z^{\kappa}} G+\ldots\ri)+ \ldots\\
      &=& \frac{L^{2\kappa +x}}{4^{\kappa} t^{\kappa}} (G+\ldots)+\ldots\;.
\ea
Since this has to be independent of $L$, it follows that
\be
\kappa\;=\;-\frac{x}{2}
\ee
so that
\be
c(t)\;\simeq\;\frac{1}{t^{-x/2}}\;=\;\frac{1}{t^{\frac{1}{2}}}\;.
\ee
This proves $\alpha=\frac{x}{2}$.

In conclusion, we can say that the study of the coagulation model gives
another example for
a successful formulation of a finite-size scaling theory for reaction-diffusion
processes, thus supporting the hypothesis that it should be possible also
for non-integrable models to extrapolate from finite to
infinite systems.
%
%----------------------------------------------------------
%
\section{Finite-Size Scaling of the Annihilation Model}
\label{sec:anni}
\hspace{\parindent}In
this section we will discuss finite-size scaling behaviour of the
concentration
of the annihilation model where we choose the rates as follows:
\ba
\mbox{diffusion rate}\;\wii(1,0) & = & \wii(0,1)\;=\;1\;,\nn\\
\mbox{annihilation rate}\;\wii(0,0) & = & 2\;,\nn\\
\mbox{creation rate}\;\wii(1,1) & = & 0\;.\nn
\ea
The Hamiltonian is then given by Eqs. (\ref{eq:HCR=0}) and (\ref{eq:H1CR0})
with $\D'=0$.
In Section \ref{sec:Traexval} we have seen that the concentration of the
annihilation and the coagulation model are related.
For homogeneous initial occupation probability $p$ the relation
(\ref{eq:konztra}) reduces to:
\be
c_{\mbox{\scriptsize{ann.}}}(p,t)=
\half c_{\mbox{\scriptsize{coag.}}}(2p,t)\,.
\ee
{}From the expression (\ref{eq:c}) for concentration in the coagulation
model we easily get:
\ba
c_{\mbox{\scriptsize{ann.}}}(t) & = & \frac{1-(1-2p)^L}{2L} \nn\\[2mm]
& &  -\;\frac{1}{2L}\;\sum_{k=1}^{L-1}\biggl[
\biggl(\frac{1+(-1)^{k+1}(1-2p)^L}{\frac{1+(1-2p)^{2}}{2(1-2p)})-
\cos{\frac{\pi k}{L}}}-\frac{1+(-1)^{k+1}(1-2p)^L}{1-\cos{\frac{\pi k}{L}}}
\biggr)\nn\\[2mm]
& &  \sin^{2}{\biggl(\frac{\pi k}{L}\biggr)}\exp
{\biggl(-4t\biggl(1-\cos{\frac{\pi k}{L}}\biggr)\biggr)}\biggr]\;.
\label{eq:ankonz}
\ea
In the finite-size
scaling limit  $L\rightarrow\inf$. Therefore the term $(1-2p)^L$
can by neglected if $0\leq p<1$. In the special case of an initially fully
occupied lattice ($p=1$)
this term has to be taken into account.
The concentration then takes the form:
\be
c(t)=\frac{1-(-1)^L}{2L}+\frac{2}{L}\sum_{k=1}^{L-1}
\frac{1-(-1)^{k+L}}{2}
\exp{\biggl(-4t\biggl(1-\cos{\frac{\pi k}{L}}\biggr)\biggr)}\;.
\label{eq:p1konz}
\ee
For $L$ even this result was first found by Lushnikov
\cite{Lushnikov}.
The conservation of
the charge $Q=(-1)^N$ is the reason for the special behaviour of the
concentration in the case $p=1$. The Hamiltonian splits into two sectors
describing the
time evolution of states with an even or odd number of particles respectively.
The momenta $q$ appearing in these sectors can be taken from the
diagonalisation in terms
of free fermions with energies given by Eq. (\ref{eq:q}).
For $p=1$ we have
$Q=(-1)^L$. Looking at Eq. (\ref{eq:p1konz}) we see that the
term $(-1)^{k+L}$ always selects the right values of the momenta $q$.
\subsection{Finite-Size Scaling of the Concentration}
\hspace{\parindent}As
in Section \ref{sec:fin_size} we calculate the finite-size scaling
behaviour by
taking the limit $L\rightarrow\inf$, $t\rightarrow\inf$ with
$z=\frac{4t}{L^2}$ fixed.
For $0\leq p<1$ the calculations can be performed in complete analogy to
those for the coagulation model. The finite-size scaling
expansion for the concentration reads:
\ba
Lc(z,L)
& = & \half\th_3(0,\frac{i\pi z}{2})\nn\\
&   &+\frac{1}{L^2}\,\half\biggl(\frac{z}{6}\frac{\partial^2}{\partial z^2}
        \th_3\biggl(0,\frac{i\pi z}{2}\biggr)
       +\frac{(1-p)^2}{2p^2}\frac{\partial}{\partial z}
\th_3\biggl(0,\frac{i\pi z}{2}\biggr)\;\biggr)
+O\biggl(\frac{1}{L^4}\biggr)\;.
\ea
As mentioned above the case $p=1$ needs a special treatment. Here we
have to discuss the finite-size scaling limit for $L$ even and $L$ odd
separately.
\begin{itemize}
\item
$L$ even:\\
In this case the concentration takes the form
\be
c(t,L)=\frac{2}{L} \sum_{k=1}^{\frac{L}{2}}
\exp{\biggl(-4t\biggl(1-\cos{\frac{2k-1}{L}\pi}\biggr)\biggr)}\;.
\label{eq:gkonz}
\ee
The calculation of the finite-size scaling limit yields
\be
Lc(z,L)=\th_2(0,2i\pi z)+\frac{1}{L^2}\:\frac{z}{3!}\:
\frac{\partial^2}{\partial z^2}\:\th_2(0,2\pi zi)
+O\left(\frac{1}{L^4}\right)
\label{eq:gkorr}
\ee
where the Jacobi theta function $\th_2$ is defined by
\be
\th_2(u,\t)=\sum_{l=-\inf}^{\inf}e^{i\pi\t(l+\half)^2}e^{2iu(l+\half)}\;.
\label{theta2}
\ee
Up to the correction term this result was already obtained by Alcaraz et al.
\cite{Alcaraz}.
\item
$L$ odd:\\
Here we have
\be
c(t,L)=\frac{1}{L}+\frac{2}{L} \sum_{k=1}^{\frac{L-1}{2}}
\exp{\biggl(-4t\biggl(1-\cos{\frac{2k}{L}}\biggr)\biggr)}\;.
\label{eq:ukonz}
\ee
and the finite-size scaling limit yields
\be
Lc(z,L)=\th_3(0,2i\pi z)+\frac{1}{L^2}\:\frac{z}{3!}\:
\frac{\partial ^2}{\partial z^2}\:\th_3(0,2\pi zi)
+O\left(\frac{1}{L^4}\right)\;.
\label{eq:ukorr}
\ee
\end{itemize}
Independently of the initial probability $p$ we have the
scaling exponent $x=-1$ and the correction exponent $y=2$ similar to the
coagulation model. This is not surprising because these models are
related by a similarity transformation.
The scaling function is again found to be independent of the initial
probability $p$ except for $p=1$. This case was already discussed.
%
%-------------------------------------------------
%
\section{Conclusions}
\label{sec:conclu}
\hspace{\parindent}In
this paper, we investigated two chemical models, the
coagulation-decoagulation model and the annihilation model, both
defined on an one-dimensional lattice. Mapping the master equation
on an euclidean
Schr\"odinger equation, it is possible to give a Hamilton
formulation for non-equilibrium chemical systems. On the one hand, we
studied the relations between the Hamiltonians of both systems, on
the other hand, we investigated the applicability of a finite-size
scaling theory for these systems.

We were able to prove the equivalence of the two Hamiltonians for a
special choice of the rates by constructing a similarity transformation
mapping one model onto the other. The consequences of this transformation,
which is valid for periodic as well as for open boundary conditions,
are various:
Once having solved one model, expectation values of the
second model can easily be found by application of the transformation.
This allows to derive all the $n$-point functions $G^{\,n}$ for one model
from the
ones calculated for the other (cf. Eq. (\ref{eq:Corrtra})).

The study of the Hamiltonians revealed furthermore that for a very special
choice of the rates, both Hamiltonians are integrable in terms of free fermions
whose energies have been calculated explicitly.

The Hamilton formalism can be used as well to calculate the mean value
of the number of particles per site
(concentration) and its time evolution. This leads to a system of differential
equations for the empty interval probabilities $\O(j,n,t)$.
{}From the solution of these differential equations,
we obtained an exact expression for the concentration in the coagulation model
where the sites are initially occupied with probability $p$.
By application of the similarity transformation, the corresponding expression
for the annihilation model was calculated easily. Observe that the
transformation lead to
values of $p$ between $0$ and $2$ in the concentration of the coagulation
model in order to obtain the concentration of the annihilation model
with $0 \leq p \leq 1$. This does not cause any problems since the
differential equations are valid for
arbitrary values of $p$.

It is very interesting that the energies determining the time-evolution of
the concentration can be identified as two-fermionic excitations of the
corresponding Hamiltonian. The question whether and how a link between
the Hamiltonian and the relevant energies for the concentration can be
established still has to be answered.

Another point of major interest was to examine the applicability
of a finite-size scaling theory for reaction-diffusion models.
Starting from the exact expression for the concentration, we have shown that
for the coagulation model and the annihilation model
the concentration in the finite-size
scaling limit can be written as $L\:c(z,L)=F_0(z)+L^{-2}\:F(z)$,
where $F_{0}(z)$ and $F(z)$ are only functions of the scaling variable
$z=\frac{4t}{L^2}$ and can be expressed in terms of Jacobi Theta functions.
A quantitative discussion of the scaling and the correction function
will be given in
the third article \cite{Pap3}. It will be shown as well that for a
different choice
of the relation between coagulation and diffusion rate, the corrections are
of the order $\frac{1}{L}$.

The comparison of the scaling behaviour of the two models supports
the conjecture
that they are both in the same universality class: the scaling and the
correction exponents are the same for both systems. Since the similarity
transformation between these models is local (i.e. independent of $L$)
and implies a simple rule of
correspondence between the expressions for the concentration,
this result was expected.
However, it is not clear whether the equivalence of the Hamiltonians generally
implies equal exponents (for example, one could imagine two models
related by a similarity transformation which depends on the lattice length
$L$ and consequently changes the scaling exponent).

As regards the scaling function $F_{0}$, we find different expressions in
both models. So the scaling function is in general model-dependent.

Since we were interested in further properties of the scaling function, we
investigated its dependence on three parameters: the details of the
model, the initial conditions and the boundary conditions. Here we sum up the
results of the present and the next two articles.

As far as the scaling function of a particular model is concerned,
it is not altered if the details of the model are changed. In the coagulation
model, for instance, it is found to be independent
of the ratio between the diffusion and the coagulation rate as we will
show using Monte Carlo simulations \cite{Pap3}, but the
corrections are altered by different tuning of the rates.

Concerning the dependence of the scaling functions on uncorrelated
initial conditions, we proved
for the coagulation model and the annihilation model that the
scaling functions are independent of the initial occupation probability.
An exception is the case of an initially completely
occupied lattice in the annihilation model where we get different
scaling functions for even and odd
lattices. The reason is the $Z_2$-symmetry. Because
the annihilation reaction reduces the number of particles by two,
even and odd lattices decouple completely.
The influence of correlated initial conditions will be investigated
in \cite{Pap3}.

Another point of interest is the dependence of the scaling behaviour
on the boundary conditions.
Observe that we used a Fourier transformation in order to solve the
system of differential equations for the $\O(j,n,t)$. This is no longer
possible using open boundary conditions. Nevertheless a solution is
possible as will be shown in our second paper \cite{Pap2}.
We will demonstrate that the exponents
are still the same; the scaling functions, however, are different for periodic
and for open boundary conditions.

We have shown the complete equivalence of the lattice and the
continuous formulation of the coagulation model. Therefore the question arises
whether it is at all necessary to study lattice models.
The previous paragraphs already contain the answer.
Monte Carlo simulations can only be made on
lattices. Furthermore, as we will show in our third article \cite{Pap3}, it is
possible to extrapolate the scaling exponent, the scaling function and the
corrections with a very high precision already from small lattices which
allows to determine the scaling behaviour of non-exactly solvable models
as well.

A lot of work is still to be done in this field. The present article shows
that the formulation of universality classes in non-equilibrium thermodynamics
is not the same as in equilibrium statistical mechanics. The annihilation
and the coagulation model are equivalent although only the annihilation
model is $Z_2$-symmetric.
A further interesting question arises looking at the scaling functions
and the correction functions:
Are they always related to Jacobi Theta functions?
The solutions of these problems will help us to gain deeper insight into
the fascinating physics of the chemical models in non-equilibrium.

\vspace{1cm}
\noindent{\large \bf Acknowledgements}
\\*[5mm]
\indent
We would like to thank Prof. V. Rittenberg for getting us into this problem
and for
enlightening discussions. We are grateful to Silvio Dahmen,
Thomas Heinzel and Gunter Sch\"utz for
critically reading the manuscript and for helpful comments.
%
%----------------------------------------------
%
\appendix
% APPENDIX: Proof of the similarity transformation
%
\section{Proof of the Similarity Transformation}
\label{sec:simpro}
\hspace{\parindent}
The Hamiltonians for the coagulation and the annihilation model
can be taken from Sec.~\ref{sec:connex}:
\ba
H_{\mbox{\scriptsize ann.}} & = & -\sum_{i=1}^{L}\biggl[
        \sp_i\sm_{i+1}+\sm_i\sp_{i+1}
        +\half\D'\sz_i\sz_{i+1}
        +\half(1-\D')(\sz_i+\sz_{i+1})
        -1+\half\D'\nn\\
 & &    +(2-2\D')\sp_i\sp_{i+1}\biggr]\;,\\
H_{\mbox{\scriptsize coag.}} & = & -\sum_{i=1}^{L}\biggl[
        \sp_i\sm_{i+1}+\sm_i\sp_{i+1}
       +\half\D'\sz_i\sz_{i+1}
       +\half(1-\D')(\sz_i+\sz_{i+1})
       -1+\half\D'\nn\\
 & &   +\half(1-\D')(\sp_i+\sp_{i+1})
       -\half(1-\D')(\sp_i\sz_{i+1}+\sz_i\sp_{i+1})
        \biggr]\;.
\ea
We will present only the proof for $b_1$. The calculations for $b_2$ can
be performed in a completely analogous way. We have:
\be
B = b^{\ten L} = \le( \begin{array}{cc}
         1 & -1\\
         0 &  2
    \end{array} \ri)^{\ten L}
\;,\hspace{1cm}
B^{-1}=(b^{-1})^{\ten L} = \le( \begin{array}{cc}
         1 & \half\\
         0 & \half
    \end{array} \ri)^{\ten L}\;.
\ee
In addition we need the following identities:
\ba
b\sp b^{-1} & = & \half\sp\;,\\
b\sm b^{-1} & = & -\sz+2\sm-\half\sp\;,\\
b\sz b^{-1} & = & \sz+\sp\;.
\ea
The application of the similarity transformation to the Hamiltonian
$H_{\mbox{\scriptsize ann.}}$ yields:
\ba
BH_{\mbox{\scriptsize ann.}}B^{-1} & = & -\sum_{i=1}^{L}\biggl[
        (b\sp b^{-1})_i(b\sm b^{-1})_{i+1}
       +(b\sm b^{-1})_i(b\sp b^{-1})_{i+1}\nn\\
  & &  +\half\D'(b\sz b^{-1})_i(b\sz b^{-1})_{i+1}
       +\half(1-\D')((b\sz b^{-1}))_i+(b\sz b^{-1})_{i+1})\nn\\
  & &  +1-\half\D'+(2-2\D')(b\sp b^{-1})_i(b\sp b^{-1})_{i+1}\biggr]\\
 & = & -\sum_{i=1}^{L}\biggl[
       -\half\sp_i\le(\sz-2\sm+\half\sp\ri)_{i+1}
       -\le(\sz-2\sm+\half\sp\ri)_i\sp_{i+1}\nn\\
  & &  +\half\D'(\sz+\sp)_i(\sz+\sp)_{i+1}
       +\half(1-\D')(\sz_i+\sz_{i+1}+\sp_i+\sp_{i+1})\nn\\
  & &  -1+\half\D'+\half(1-\D')\sp_i\sp_{i+1}\biggr]\\
 & = & -\sum_{i=1}^{L}\biggl[
        \sp_i\sm_{i+1}+\sm_i\sp_{i+1}
       +\half\D'\sz_i\sz_{i+1}
       +\half(1-\D')(\sz_i+\sz_{i+1})-1+\half\D'\nn\\
  & &  +\half(1-\D')(\sp_i+\sp_{i+1})-\half(\sp_i\sz_{i+1}+\sz_i\sp_{i+1})
       \biggr]\\
 & = & H_{\mbox{\scriptsize coag.}}\;.
\ea
This completes the proof.
%
%------------------------------------------
%
% Appendix ueber Kontinuum-Formulierung
%
\section{Derivation of the Scaling Function
\protect\\
in the Continuum Limit}
\label{sec:Cont}
\hspace{\parindent}
In this appendix, we present the continuum limit of the differential
equations (\ref{eq:hol1}) for periodic boundary conditions, treating the
coagulation-decoagulation model with diffusion rate $=$ coagulation rate
and decoagulation rate $= 0$ (i.e. $\e=1$).
The continuum limit is obtained by replacing the discrete variable $n$,
which denotes the length of an empty interval in Sec.~\ref{sec:holes}
by a continuous variable $x\;\epsilon \;[0,L]$. Because of periodic boundary
conditions, the model is defined on a ring of circumference $L$.
We define a new function $E(x,t)$ describing  the
probability that a chosen interval of length $x$ is empty at time $t$.
Because we are only interested in observables that are averaged over the
whole system
we define $E(x,t)$ independently of the center-point.
So $E(x,t)$ corresponds to the continuum limit of $\Oh(0,n,t)$ (the Fourier
transformation of the $\O(j,n,t)$ with respect to momentum $0$ of
Sec.~\ref{sec:holes}).
Following the derivation given by \cite{ben-Avraham2}, we obtain the
differential equation for $E(x,t)$:
\be
\frac{\partial E(x,t)}{\partial t} \;=\;  2 \;\frac{\partial^{2}E(x,t)}
{\partial x^{2}}.
\ee
This is the well known heat equation.
Before solving the differential equation, we first change the variable $x$
to $y=\frac{x}{L}$ so that $y\;\epsilon \;[0,1]$.
$E(x,t)$ changes to:
\be
E(x,t)\;=\;\tilde{E}(y,t)
\ee
and the differential equation becomes:
\be
\frac{\partial \Et(y,t)}{\partial t} \;=\; \frac{2}{L^2}\frac
{\partial^{2}\tilde{E}(y,t)}{\partial y^{2}}
\label{eq:heat}
\ee
with the boundary conditions:
\ba
\Et(0,t)&=& 1\;,\nn\\
\Et(1,t)&=& \Et(1,0).
\label{eq:bound}
\ea
The concentration can be calculated as a derivative of $\tilde{E}(y,t)$:
\be
c(t)\;=\;\le.-\frac{1}{L}\frac{\pa\tilde{E}(y,t)}{\pa y}\ri\vert_{y=0}\;.
\label{eq:deri}
\ee
We now briefly present the solution of Eq. (\ref{eq:heat}).
As initial condition we choose:
\be
\Et(y,0)\;=\;\exp(-c_0 L y).
\ee
This probability distribution corresponds to an uncorrelated initial state
with an initial concentration $c_0$.
A particular solution (steady state) satisfying the boundary conditions
(\ref{eq:bound})
is given by:
\be
\Et_s(y)\;=\;1-\le[1-\exp\le(-c_0 L\ri)\ri]\;y\;.
\ee
First solving the homogeneous problem, we observe that $\Et(y,t)$
can be expanded in terms of the eigenfunctions $\psi_k(y)$
of the eigenvalue problem:
\be
\l_k\;\psi_k(y)\;=\;\frac{2}{L^2} \frac{\partial^{2}\psi_k(y)}{\partial y^2}
\ee
with:
\ba
\psi_k(0)& = & 0\;,\\
\psi_k(1)& = & 0.
\ea
Then $\Et(y,t)$ can be expressed in terms of the eigenfunctions as:
\be
\Et(y,t)\;=\;\sum_k\; B_k \psi_k(y) e^{\l_kt}.
\ee
The solution of this problem is easily determined to be:
\ba
\psi_k(y) &=& \sin \le(\pi k y\ri)\;,\\
\l_k &=& -\frac{2 \pi^2 k^2}{L^2}\;,\\
k &=& 0,1,2,\ldots\;.
\ea
The eigenfunctions $\psi_k(y)$ satisfy the orthogonality relation:
\be
2\;\int_0^1 \sin(\pi k y)\;\sin (\pi l y) dy \;=\;\delta_{k,l}\;.
\label{eq:ortho}
\ee
\\
The complete solution of the differential equation (\ref{eq:heat}) is
now given by:
\be
\Et(y,t)\;=\;\Et_s(y)\;+\;\sum_{k=0}^{\infty}\; B_k \sqrt{2}\sin(\pi k y)
e^{-\frac{2 \pi^2 k^2 t}{L^2}}
\ee
where the coefficients $B_k$ are calculated with the help of the orthogonality
relation (\ref{eq:ortho}):
\ba
B_k&=&\int_0^1 \le[\Et(y,0)-\Et_s(y)\ri]\sqrt{2}\sin(\pi k y) dy\\
   &=&\sqrt{2}\le[\frac{1}{1+\le(\frac{\pi k}{c_0 L}\ri)^{2} }\ri]
              \le[\frac{\exp(-c_0 L)\cos(\pi k)-1}{\pi k}\ri].
\ea
Now we are able to derive the concentration using Eq. (\ref{eq:deri}) as:
\ba
c(t,L)& =& \frac{1-\exp(-c_0 L)}{L}\nn\\
      &  &     +\frac{2}{L}\sum_{k=0}^{\infty}
          \le[1-\exp(-c_0 L)\cos(\pi k)\ri]
           \le[\frac{1}{1+\le(\frac{\pi k}{c_0 L}\ri)^{2} }\ri]
          \exp\le(-\frac{2t\pi^2 k^2}{L^2}\ri).
\ea
This expression is exact for all lengths $L$.

Let us now turn to the finite-size scaling limit $L\rightarrow \infty$,
$t\rightarrow\infty$ but $z=\frac{4 t}{L^2}$ fixed.
In this limit, the above expression for the concentration becomes:
\ba
c(z,L)& =& \frac{1}{L}
          +\frac{2}{L}\sum_{k=1}^{\infty}
                      \exp\biggl(-\frac{z \pi^2 k^2}{2}\biggr)
          -\frac{2}{L^3 c_0^2}
           \sum_{k=1}^{\infty}\pi^{2} k^{2}
               \exp\biggl(-\frac{z \pi^2 k^2}{2}\biggr)
           +  \ldots\\
       & =& \frac{1}{L}\le[\theta_3\biggl(0,\frac{i \pi z}{2}\biggr)
            +\frac{2}{L^2 c_0^{2}}\,
           \frac{ \partial}{\partial z}\,\theta_3\biggl(0,
                 \frac{i \pi z}{2}\biggr)
            +\ldots\ri]\;.
\label{eq:scalcont}
\ea
{}From the expansion of this expression for small $z$ we can calculate
the long-time behaviour in the thermodynamic limit:
\be
c(t) = \frac{1}{\sqrt{2\pi t}}\le[1-\frac{1}{4tc_0^2}+\cdots\ri]\;.
\label{eq:ltimecont}
\ee
We now compare Eqs. (\ref{eq:scalcont}) and (\ref{eq:ltimecont}) with
the corresponding expressions Eqs. (\ref{eq:scf}) and (\ref{eq:ltime})
for the lattice model. The leading terms are equal but the corrections
differ. In order to understand this, we take a closer look on the
way the continuum limit of the lattice model has to be taken.
Here we denote by $N$ the number of sites to distinguish it from
the length $L$ of the continuous model, $a$ is the lattice constant, i.e.
the distance between to sites. Taking the continuum limit means:
\be
N \lra \infty \hspace{1cm} \mbox{and} \hspace{1cm} a \lra 0
\ee
while several constraints have to be satiesfied:
\begin{itemize}
\item the length $L=Na$,
\item the macroscopic diffusion constant $D=a^2\wii(1,0)$
\item and the initial concentration $c_0 = p/a$
\end{itemize}
have to remain constant.
Therefore the transition from the lattice to the continuum is performed
by replacing:
\ba
c         & \lra & \frac{c}{a}=\frac{c\,c_0}{p}\;,\\
N         & \lra & \frac{L}{a} = \frac{Lc_0}{p}\;,\\
\wii(1,0) & \lra & \frac{D}{a^2} = \frac{Dc_0^2}{p^2}
\ea
and then taking the limit:
\be
p \lra 0\;.
\ee
It is clear that the value of the scaling variable $z$ which is
$z=\frac{4\wii(1,0)t}{N^2}$ on the
lattice and $z=\frac{4Dt}{L^2}$ in the continuum is not affected.

If we apply the above procedure to Eqs. (\ref{eq:scf})
and (\ref{eq:ltime}) for the lattice model, we obtain exactly
Eqs. (\ref{eq:scalcont}) and (\ref{eq:ltimecont})
for the continuous model.

To sum up, we have shown that it is
possible to turn to the continuum limit of the coagulation
model to study the scaling behaviour. This should be
possible for all models with reactions that can be described in the language
of empty interval probabilities where the calculation of the continuum limit
is straightforward. The physics of the continuum model is the same as
for the discrete model
but the differential equations are often much easier to solve.
%
%------------------------------------------------
%

\end{document}